# Theory and New Amplification Regime in Periodic Multi Modal Slow Wave Structures with Degeneracy Interacting with an Electron Beam

Mohamed A. K. Othman, Venkata Ananth Tamma and Filippo Capolino

*Department of Electrical Engineering and Computer Science, University of California, Irvine, CA, USA 92697.*

*Abstract* —We present the theory of a new amplification regime in Travelling Wave Tubes (TWTs) composed of a slow-wave periodic structure that supports multiple electromagnetic modes that can all be synchronized with the electron beam. The interaction between the multimodal slow-wave structure and the electron beam is quantified using a Multi Transmission Line approach (MTL) based on a generalized Pierce model and transfer matrix analysis leading to the identification of modes with complex Bloch wavenumber. In particular, we address a new possible operation condition for TWTs based on the super synchronism between an electron beam and four modes exhibiting a degeneracy condition near a band edge of the periodic slow-wave MTL. We show a phenomenological change in the band structure of periodic MTL where we observe at least two growing modal cooperating solutions as opposed to a uniform MTL interacting with an electron beam where there is rigorously only one growing modal solution. We discuss the advantage of using such a degeneracy condition in TWTs that leads to larger gain conditions in amplifier regimes and also to very low-starting beam current in high power oscillators.

## I. INTRODUCTION

High power microwave (HPM) devices have many applications in diverse areas such as RADAR, satellite and terrestrial communications as well as electromagnetic interference and jamming [1], [2]. Of the many different types of HPM devices, the travelling wave tube (TWT) is very widely used due to its simplicity, low cost, reliability and wide bandwidth [3]–[6]. An important component of the TWT is the slow-wave structure (SWS), where the electron beam interacts with a slow electromagnetic wave leading to energy transfer from the beam to the radio frequency (RF) wave over some distance [7]–[9]. The energy transfer is based on a mode with exponential growth along the length of the TWT so that the amplitude of the input electromagnetic wave is amplified at the TWT output. The choice of the SWS is critical for best operation of the TWT. The helix is a popular SWS that has been used due to the its simplicity of construction, large interaction impedance and wide operating bandwidth [10]–[12]. Apart from the helix, periodic structures mainly made up of coupled cavities or slots have previously been utilized as SWS in TWTs [8], [13], [14]. Some other commonly used SWSs include comb structure [15], disk structures [16], interdigitated structures[17] and folded-waveguides for terahertz radiation [18], [19], among others. Metamaterial based slow-wave structures were proposed recently [20] by embedding subwavelength resonant elements inside waveguides, like ring-bars [21] or split-ring resonators [22], [23]. Typically at microwave frequencies, periodic structures are implemented as metallic obstacles or slots due to lower losses than in dielectrics and to the ability to sustain very large field gradients [24]. Such periodic SWS are usually operated at frequencies far from the edge of their first Brillouin zone so as to achieve synchronism with the electron beam for amplification regimes. Backward modes in a periodic SWS are usually utilized for backward wave amplifiers and oscillators [25], [26].

The underlying physical operation of the TWT with SWS is well described by resorting to equivalent transmission lines (TLs) with the charge wave supported by the electron beam modelled as a charge fluid, according to a model introduced by Pierce [4], [5]. Following Pierce model, the amplification in a TWT is readily attributed to amplification of a slow RF signal in an equivalent TL due to charge density oscillations (the charge wave) of the electron beam. Despite the abstraction of the SWS as a TL and the hydrodynamic modelling of the electron beam, the Pierce model has proven to be robust and reasonably accurate engineering tool to predict TWT's behaviour and performance under small-signal approximation. A more elaborate theory of interaction, such as non-stationary nonlinear formalism of TWTs was developed in [27]–[29] particularly to study the effect of operation very close to the edge of the Brillouin zone at a regular band edge (or equivalently, near the cutoff frequency using the language in those papers), where the transition between amplification to self-excitation and oscillation regimes was reported in finite length structures. Typically, periodic SWS waveguides support several Bloch modes (propagating and evanescent), where each Bloch mode possesses an infinite number of spatial Floquet harmonics of which only one of the spatial harmonic is synchronized with the electron beam. If that particular synchronized spatial harmonic is amplified, all the other harmonics of the relevant mode would be amplified, as typical in periodic structures.

Here, we investigate an amplification method based on the significant reduction in the group velocity of electromagnetic modes in a periodic SWS generated by a degeneracy of four slow-wave Bloch modes. The idea is based on work by Figotin and Vitebskiy [30]–[35] that have proposed frozen mode regimes in multilayer dielectric structures. This regime generates a dramatic increase in field intensity associated with a transmission band-edge "giant" resonance achieved by repeating unit cells made of specially arranged periodic combinations of anisotropic dielectric layers with a misalignment of in-plane anisotropy. An interesting property of this novel band-edge resonance, exploited in this paper, is a degenerate band edge (DBE) condition that causes a quartic power dependence at the band-edge of the Bloch wavenumber $k$, i.e., $\Delta\omega \propto (\Delta k)^4$. Moreover, not only the group velocity vanishes at the DBE frequency, i.e., $v_g = \partial_k\omega = 0$, but also the first and second derivative of the group velocity are identically zero, i.e., $\partial_k v_g = \partial_k^2\omega = 0$ and $\partial_k^2 v_g = \partial_k^3\omega = 0$, with the only the third derivative of the group velocity non-zero, $\partial_k^3 v_g = \partial_k^4\omega \neq 0$ (where the partial derivative is defined as $\partial_k^n\omega \equiv \partial^n\omega / \partial k^n$). This implies a gigantic increase in the density of electromagnetic modes [36]. Due to the extremely low group velocity in the vicinity of the DBE and large field enhancement, structures exhibiting DBE modes are here explored for their potential use as novel SWSs in TWTs for high power generation. Analogous considerations apply for other degenerate conditions like the split band edge, investigated in [34], [35], [37], that we will consider in future studies. At the RF frequency range, the DBE condition has been already observed in periodically-coupled microstrip lines [38], [39]. It has been observed also at optical frequencies in coupled optical silicon waveguides [40] and we have observed it also in circular waveguide structures like those in Fig. 1, and as those in [41]. The MTL model described here is suitable to describe the DBE in all these configurations, including other geometries like those in Fig. 1.

In previous work, we have extended the "one-dimensional" Pierce model to the study of the interaction of MTLs with an electron beam [42]. A formalism was developed to calculate the $k$-$\omega$ dispersion relation and conditions required to achieve growing waves in the system were discussed, including an equivalent circuit model to find the complex modes. In this paper, we use the transfer matrix formulation presented in [42] to study a *periodic* MTL structure coupled to a single electron beam. Similar to our previous work, for simpler analysis we assume a small signal approximation for the charge wave [42]. There, two examples of periodic SWS waveguides that are able to support a DBE condition are shown, where a unit cell is constructed (a) by cascading three different waveguide cross sections with different ellipticity (aspect ratios of the elliptical cross-sections), or (b) by loading the waveguide with metallic rings of elliptic shapes that can be rotated by a misalignment angle $\delta\varphi$ between the major axes of the rings and varied in thickness. A particular design of those structures supporting a DBE were reported by the authors in [41] where the dispersion diagram of the waveguide is obtained using full wave simulations. All these structures, and many more [38], [43], can be conveniently modeled using the MTL formalism (Fig. 2). The theory shown here, generalizing the Pierce model as in [42], is used to study the interaction of an electron beam with periodic structures with dispersive modes that exhibit a DBE condition. Moreover, the SWS-beam interaction very close the degenerate band edge has not been explored yet in the literature; only the SWS–beam interaction near a regular band edge (RBE) has been investigated in a few studies such as those in Refs. [26]–[29].

In Sec. II, we derive the transfer matrix formalism for a generalized *periodic* MTL structure assuming linear time-harmonic fields varying as $e^{j\omega t}$ . In Secs. III-IV, we apply the transfer matrix formalism derived in Sec. II to study the effect of the electron beam on the dispersion diagrams. In particular, we are interested in the particular case when the four degenerate electromagnetic modes operating in proximity of the DBE condition, are simultaneously synchronized with the electron beam. A condition we refer to as *super synchronism,* when all four electromagnetic Bloch modes in principle interact with the electron beam (to be precise a Floquet harmonic of each Bloch mode interacts with the electron beam). We also show how the interaction with the electron beam slightly disrupts the degeneracy condition. We quantitatively analyze modal propagation in the periodic MTL interacting with an electron beam and determine the six modal solutions described by complex-valued Bloch wavenumbers, which describe propagation along positive and negative $z$ directions. We show that the degree of the DBE perturbation depends on the amplitude of the electron beam average current. The super synchronous condition between the beam and the four degenerate modes causes the manifestation of more than one mode with growing amplitude along the $z$-direction. This is in contrast to homogeneous MTL-electron beam systems that support identically only one growing modal solution, as shown in [42] and rigorously proved in [44].  Hence, the super-synchronous condition may have a possible impact on future TWTs working with feeble electron beams. Relative to the concept of super synchronism, we also show the weight of each Floquet harmonic in the periodic MTL system and show that the higher order Floquet harmonics of the electron beam could be neglected in this very special super synchronous condition. In this study we mainly analyze periodic-MTL systems with infinite length. In Sec. V, a case of MTL of finite length, i.e., a MTL system with boundary conditions, with excitation and loading is shown to exhibit larger gain compared to the "one-dimensional" Pierce model and we identify some advantages of using our DBE structure to enhance gain in TWTs.

## II. TRANSFER MATRIX FORMALISM FOR THE PERIODIC MTL SYSTEM INTERACTING WITH AN ELECTRON BEAM

In this section, we follow the Pierce model using the transfer matrix formalism for MTL interacting with an electron beam derived in [42] which is extended to the case of MTLs consisting of *periodic* sets of $M$ cascaded MTL segments $A_1, A_2,.., A_M$ shown in Fig. 2(a). Each segment is assumed to be a MTL consisting of $N$ TLs, coupled among each other by capacitive and/or inductive means. In each segment, each TL may be coupled to the electron beam with distinct coupling coefficients [42]. The electron beam is assumed to propagate along the $+z$ direction. The lengths of the segments $A_1, A_2,.., A_M$ are $d_1, d_2,.., d_M$ and the total periodic unit cell length is $d = \sum_{m=1}^{M} d_m$ . We assume that the $l^{th}$ TL, where $l = 1, 2,\ldots, N$ , in the $m^{th}$ segment, with $m = 1, 2,\ldots, M$ , interacts with the beam described by the field interaction coefficient $a_{m,n}$ and the current interaction factor $s_{m,n}$ , where the field and current interaction factors were defined in [42]. Weights of electric and magnetic fields are equivalently represented using the vectors $\mathbf{V}(z) = [V_1, V_2,.., V_N]^T$ and $\mathbf{I}(z) = [I_1, I_2,.., I_N]^T$ , respectively, as we show in Appendix A. The field weights are traditionally defined as equivalent TL voltages and currents [45], [46]. The electron beam is described as in [42], according to the Pierce model [1]–[5], with an average beam current $I_0 = \rho_0 u_0$ , where $\rho_0$ is the average beam charge per unit length, $u_0$ is the average (d.c.) electron velocity, and $\beta_0 = \omega / u_0$ is the phase constant of a wave travelling at the electrons stream average velocity $u_0$ [5]. The time-harmonic varying (or a.c.) beam current is represented by the phasor $I_b$ that represents a charge wave travelling with speed $v_b$ (also a phasor, since it is also time-harmonic, oscillating with the same frequency of the electromagnetic wave). It is convenient to define $V_b = u_0 v_b / \eta$ as an equivalent kinetic beam voltage (a.c. modulation or "charge wave" voltage) with units of [V], as done in [42], [47], where $\eta = e / m = 1.759 \times 10^{11}$ [C/kg] denotes the charge-to-mass ratio of the electron with charge equal to $-e$ and $m$ is the electron mass.

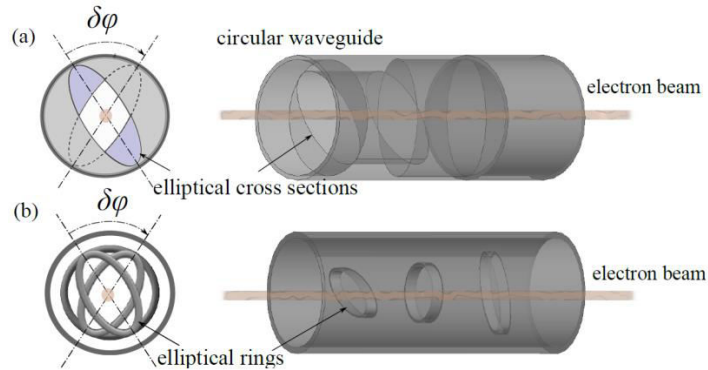

Fig. 1. Schematic of waveguide unit cells as constituents of periodic structures along the *z* direction that may support several Bloch modes and in particular a degenerate band edge. In (a) The unit cell is composed of a few waveguide cross-section with different ellipticities or orientations, whereas in (b) the waveguide is loaded with a few rings with different ellipticities or orientations.

We then define a state vector as $\boldsymbol{\Psi}(z) = \begin{bmatrix} \mathbf{V}^T(z) & \mathbf{I}^T(z) & V_b(z) & I_b(z) \end{bmatrix}^T$ with dimension $(2N+2)$ that describes propagation along $z$ of fields and linear charge wave in the MTL-beam system, and satisfies

$$\partial_z \boldsymbol{\Psi}(z) = -j\underline{\mathbf{M}}(z)\boldsymbol{\Psi}(z) \,, \tag{1}$$

where $\underline{\mathbf{M}}$ is a system matrix that describes all the $z$-dependent TL and electron beam parameters, including coupling effects [42]. In each $m^{th}$ segment shown in Fig. 2, we assume the TL parameters and hence the matrix $\underline{\mathbf{M}}$ constant and given by [42]

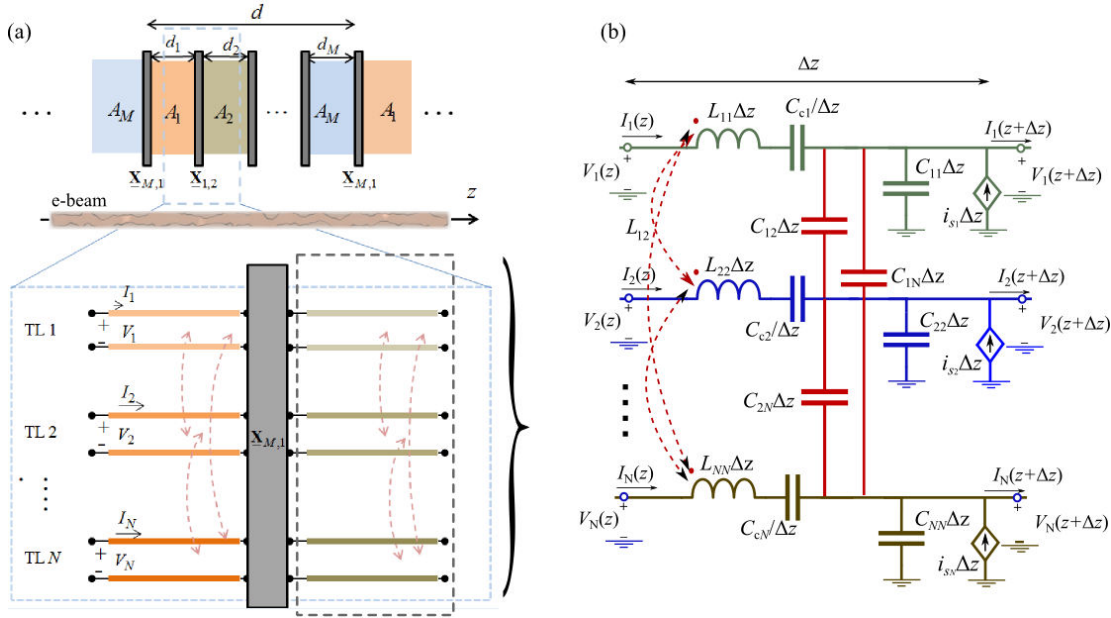

Fig. 2. (a) Schematic of an electron beam interacting with a periodic MTL system. The periodic MTL system consists of $M$ segments $A_1$, $A_2$,..., $A_M$ of coupled transmission lines, with additional coupling matrix $\underline{\mathbf{X}}$ at the interface between contiguous segments. (b) The distributed circuit model of one segment of MTL, showing self, coupling and "cutoff" components (represented by a series capacitance in each TL), as well as the electron beam seen as a dependent current generator that provides energy to the MTL.

$$
\underline{\underline{\mathbf{M}}}_m = \begin{bmatrix}
\underline{\underline{\mathbf{0}}} & -j\underline{\underline{\mathbf{Z}}}_m & \mathbf{0} & \mathbf{0} \\[2mm]
-j\underline{\underline{\mathbf{Y}}}_m & \underline{\underline{\mathbf{0}}} & \omega\eta\,\dfrac{\rho_0}{u_0^2}\mathbf{s}_m & -\beta_0\mathbf{s}_m \\[2mm]
\mathbf{0} & j\left(\mathbf{a}_m^T\underline{\underline{\mathbf{Z}}}_m\right) & \beta_0 & -\dfrac{\omega_p^2}{\omega\rho_0\eta} \\[2mm]
\mathbf{0} & \mathbf{0} & -\omega\eta\,\dfrac{\rho_0}{u_0^2} & \beta_0
\end{bmatrix},
\tag{2}
$$

where $\underline{\underline{\mathbf{Z}}}_m$ and $\underline{\underline{\mathbf{Y}}}_m$ are the $N \times N$ per-unit-length series impedance and shunt admittance matrices, respectively, of the $m^{\text{th}}$ MTL segment. For the MTL considered in this paper we have $\underline{\underline{\mathbf{Z}}}_m = \underline{\underline{\mathbf{R}}}_m + j\omega\underline{\underline{\mathbf{L}}}_m + \dfrac{1}{j\omega}\underline{\underline{\mathbf{C}}}_{c,m}^{-1}$ and $\underline{\underline{\mathbf{Y}}}_m = j\omega\underline{\underline{\mathbf{C}}}_m$, where $\underline{\underline{\mathbf{L}}}_m$ and $\underline{\underline{\mathbf{C}}}_m$ are the $N \times N$ inductance and capacitance matrices of the $m^{th}$ segment, respectively (assumed to be symmetric, and positive definite matrices [48], [49]), and $\underline{\underline{\mathbf{R}}}_m$ is $N \times N$ loss resistance matrix accounting for dissipative losses in the MTL. Moreover, for modeling a waveguide with cutoff frequency with an equivalent TL, we have to consider a cutoff condition for each of the TL modes that propagate in the $z$-direction. This is done by incorporating a series capacitance matrix $\underline{\underline{\mathbf{C}}}_c$ (each element in $\underline{\underline{\mathbf{C}}}_c$ matrix has units of [F·m] ensuring that the per-unit-length impedance has units [$\Omega$/m]) in the distributed MTL model as depicted in Fig. 2(c). This capacitance inhibits propagation of low frequencies, similarly to a cutoff condition associated with TM$^z$ waveguide modes [50], and it is assumed for simplicity to be a diagonal matrix, though the formulation here is general. Other definitions, notations, and properties of the parameters pertaining to the MTL and electron beam that are not elaborated here are found in [42]. The space charge effect ("debunching" of charge waves due to self-generated forces [5]) is also included in the system matrix. Those self-forces induce oscillations of the charge wave with the plasma frequency defined as $\omega_p^2 = n_v e^2 / (\varepsilon_0 m) = \rho_0 \eta / (A\varepsilon_0)$ where $n_v$ is the volumetric electron density, $A$ is the beam area, and $\varepsilon_0$ is the permittivity of vacuum. All the parameters values used in numerical simulations are listed in Appendix B.

A.    *Evolution of fields in the periodic MTL interacting with an electron beam*

When considering the boundary condition $\mathbf{\Psi}(z_0) = \mathbf{\Psi}_0$ at a certain coordinate $z_0$, the system (1) represents the well-known Cauchy's problem. The state vector solution at a coordinate $z_1$ is found via

$$
\mathbf{\Psi}(z_1) = \underline{\underline{\mathbf{T}}}(z_1, z_0)\,\mathbf{\Psi}(z_0),
\tag{3}
$$

where we define $\underline{\underline{\mathbf{T}}}(z_1, z_0)$ as the *transfer matrix* which uniquely relates the state vector $\mathbf{\Psi}(z)$ between two points $z_0$ and $z_1$ (we assume $z_1 > z_0$) along the $z$ axis [42]. Note that we always have $\underline{\underline{\mathbf{T}}}(z_1, z_0)\underline{\underline{\mathbf{T}}}(z_0, z_1) = \underline{\underline{\mathbf{1}}}$ with $\underline{\underline{\mathbf{1}}}$ being the identity matrix [33], [51], that has the same dimension as $\underline{\underline{\mathbf{T}}}$. If both $z_0$ and $z_1$ belong to the same homogeneous segment of TL, the transfer matrix is simply found as $\underline{\mathbf{T}}(z_1, z_0) = \exp(-j(z_1 - z_0)\underline{\mathbf{M}})$. When they belong to different segments, transfer matrix multiplication is required, as detailed later.

As shown in Fig. 2, we assume that the junction (interface) between any two adjacent segments $A_m$ and $A_{m+1}$ is generally represented by a network with $2(N+1)$ ports, and described by a transfer matrix $\underline{\mathbf{X}}_{m,m+1}$. It is assumed that it models the fields and beam transfer between two contiguous segments (for example two waveguides with different cross-sections in Fig. 1(a)), and it does not have any physical length. Therefore, the $N \times N$ sub-block MTL terms in $\underline{\mathbf{X}}_{m,m+1}$ represent equivalent lumped elements like inductors, capacitors, transformers, etc. This coupling matrix $\underline{\mathbf{X}}_{m,m+1}$ can also represent a physical component or surface that rotates the state of the polarization [52]. Though not done in this paper, a coupling matrix $\underline{\mathbf{X}}_{m,m+1}$ can describe also the interface between misaligned anisotropic dielectric layers that were considered in [34], therefore it is a general description that mixes state quantities at an interface between any two contiguous segments. A few cases are listed in Appendix A.

Furthermore, the beam's equivalent voltage and current $V_b(z)$ and $I_b(z)$ are continuous across a junction between two waveguide segments (Fig. 1), and the $2 \times 2$ block matrix representing the beam mixing term is an identity matrix. Therefore the coupling elements $\underline{\mathbf{X}}_{m,m+1}$ are described by the matrix

$$\underline{\mathbf{X}}_{m,m+1} = \begin{bmatrix} \underline{\underline{\mathbf{\Phi}}}_m^{\mathbf{VV}} & \underline{\underline{\mathbf{\Phi}}}_m^{\mathbf{VI}} & \mathbf{0} & \mathbf{0} \\ \underline{\underline{\mathbf{\Phi}}}_m^{\mathbf{IV}} & \underline{\underline{\mathbf{\Phi}}}_m^{\mathbf{II}} & \mathbf{0} & \mathbf{0} \\ \mathbf{0} & \mathbf{0} & 1 & 0 \\ \mathbf{0} & \mathbf{0} & 0 & 1 \end{bmatrix}, \tag{4}$$

which transforms the state vector $\mathbf{\Psi}(z)$ across the interface between any two adjacent segments $A_m$ and $A_{m+1}$.

In general, junctions (interfaces) between adjacent waveguide with different cross-sections (see Fig. 1) are typically associated also with higher order evanescent fields and in such a case the TL blocks in $\underline{\mathbf{X}}_{m,m+1}$ may represent reactive circuit elements. Since the goal of this paper is to show new amplification regimes, we do not refer to a specific geometry. Therefore we do not study the interaction of higher order waveguide modes, and we focus on the $N$ TL modes considered here. A simple rotation operator applied that transforms vectors $\mathbf{V}_m$, $\mathbf{I}_m$ to $\mathbf{V}_{m+1}$, $\mathbf{I}_{m+1}$ is a particular case of this more general formalism (see appendix A). This rotation operation describes mode coupling at the junction between two waveguides with rotated elliptical cross section, as in Fig. 1, for example. In this cases the terms $\underline{\underline{\mathbf{\Phi}}}^{\mathbf{VV}}, \underline{\underline{\mathbf{\Phi}}}^{\mathbf{VI}}, \underline{\underline{\mathbf{\Phi}}}^{\mathbf{IV}}, \underline{\underline{\mathbf{\Phi}}}^{\mathbf{II}}$ define projections between $\mathbf{V}$ and $\mathbf{I}$ across the interface. More details are found in Appendix A. The rotation case is discussed in Sec. III.

Now we recall the transfer matrix $\underline{\mathbf{T}}(z_1, z_0)$ in (3) and we consider a periodic MTL structure that has a unit cell of length $d$, comprising of $M$ segments as in Fig. 2(a), each of which is described by a transfer matrix $\underline{\mathbf{T}}_m = \underline{\mathbf{T}}_m(z_{m+1}, z_m)$ with $z_{m+1} - z_m = d_m$. Coupling between any two contiguous segments is portrayed by $\underline{\mathbf{X}}_{m,m+1}$. We obtain the transfer matrix of the unit cell by cascading the transfer matrices of individual segments

$$\underline{\mathbf{T}} = \underline{\mathbf{T}}_M \, \underline{\mathbf{X}}_{M-1,M} \dots \underline{\mathbf{T}}_2 \, \underline{\mathbf{X}}_{1,2} \, \underline{\mathbf{T}}_1 \, \underline{\mathbf{X}}_{M,1}, \tag{5}$$

such that the state vector evolves across the unit cell as

$$\mathbf{\Psi}(z+d) = \underline{\mathbf{T}}(z+d, z)\, \mathbf{\Psi}(z). \tag{6}$$

Considering an infinite long periodic MTL system, we seek periodic solutions of the state vector $\mathbf{\Psi}(z)$ in the Bloch form that evolve as

$$\mathbf{\Psi}(z+d) = e^{-jkd}\mathbf{\Psi}(z), \tag{7}$$

where $k$ is the complex Bloch wavenumber. Then we write the eigensystem by combining (6) and (7) as

$$\underline{\mathbf{T}}(z+d, z)\, \mathbf{\Psi}(z) = e^{-jkd}\mathbf{\Psi}(z), \tag{8}$$

whose eigenvalues are $e^{-jk_n d}$, with $n = 1, 2, .., 2(N+1)$, and are evaluated by solving the characteristic equation

$$\det\left[\underline{\mathbf{T}}(z+d, z) - e^{-jkd}\, \underline{\mathbf{1}}\right] = 0 \tag{9}$$

for complex $k$. We call $\mathbf{\Psi}_n(z)$ the $n^{\text{th}}$ eigenvector of the transfer matrix $\underline{\mathbf{T}}(z+d, z)$. Note that all the eigenvalues are not $z$-dependent, whereas the state eigenvectors depend on the coordinate $z$.

### B.    Floquet wave expansion of slow-wave TL modes and charge wave

TL voltages and currents in a periodic MTL (or equivalently electric and magnetic propagating fields in a periodic waveguide) interacting with the beam are solutions of (8). This means that both TL quantities and charge-wave are all represented as a superposition of spatial Floquet harmonics as

$$\mathbf{\Psi}_n(z) = \sum_{p=-\infty}^{\infty} \mathbf{\psi}_n^p e^{-jk_n^p z}, \tag{10}$$

where $k_n^p = k_n + F_p$ is the wavenumber of the $p^{\text{th}}$ Floquet harmonic, with $F_p = 2\pi p / d$, where $p = 0, \pm 1, \pm 2, \dots$ and $k_n$ is the fundamental Bloch wavenumber of the $n^{\text{th}}$ mode obtained from (9). The fundamental Bloch wavenumber is the one satisfying $-1 < \text{Re}(k_n d / \pi) < 1$. Note that all Floquet harmonics retain the same imaginary part, i.e., $\text{Im}(k_n^p) = \text{Im}(k_n)$, that means that if the fundamental harmonic of a mode is growing due to the beam energy transfer, all the associated Floquet harmonics are growing with the same rate. The state vector weights $\mathbf{\psi}_n^p$ are found by projection integrals over a period of the coupled TL-beam system:

$$\mathbf{\Psi}_n^p = \frac{1}{d} \int\limits_0^d \mathbf{\Psi}_n(z)\, e^{+jk_n^p z} dz \,. \tag{11}$$

In Sec. III, we show dispersion diagrams of the Block wavenumber $k$ in the first Brillouin zone. Note that due to the periodicity of the MTL also the charge wave becomes a periodic function of $z$ (besides an exponential term, therefore in the language of many, it is quasi-periodic). The good approximation in neglecting higher order Floquet harmonics for the charge wave is demonstrated in Sec. III, based on the observation that weights of higher-order Floquet harmonics of the charge wave are significantly lower than those of the fundamental harmonic (by many dBs) for the cases we investigate in this premise. Moreover, due to MTL coupling with the electron beam, the $\pm k$ symmetry of wavenumber solutions is not generally preserved (this symmetry is satisfied rigorously in reciprocal systems, i.e., in the "cold" MTL, when it does not interact with the electron beam).

The formulation outlined so far serves as a general framework for treating periodic structures that support $N$ TL modes and takes into account the coupling to an electron beam. In the next two sections we apply this formalism to study a few examples of periodic MTL systems. In particular, we study the effect of the electron beam on the dispersion characteristic of TL structures that support the DBE condition. All parameters used in the calculations are detailed in Appendix B.

### III. MODES IN PERIODIC MTL SYSTEMS: DBE CASE WITH INTERFACE MIXING

We present here an illustrative example involving a periodic structure consisting of two TLs where voltages and currents are coupled at the junctions between two contiguous TL segments as shown in Fig. 3. The coupling $\underline{\mathbf{X}}$ considered here is as in (4) and it represents a rotation matrix for TL voltages and currents. In particular, we choose an MTL structure that supports a DBE condition, when not coupled to the electron beam (i.e., a "cold" MTL) and then we investigate its interaction with the electron beam and observe how the dispersion diagram with DBE is modified. To understand the reasons of why the proposed TL exhibits the DBE condition we recall here the first investigation of structures supporting a DBE proposed by Figotin and Vitebskiy in [31], [32], [34], [35], [51]. Such periodic structures were composed of multi-layered anisotropic (birefringent) dielectric media whose unit cell consisted of two misaligned anisotropic layers and one isotropic layer. The two anisotropic layers had in-plane anisotropy with two different orientations of the optical axes, say the main axis is oriented along $\varphi_1$ and $\varphi_2$ in two contiguous layers, respectively. The angle $\delta\varphi = \varphi_1 - \varphi_2$ defines the optical axis misalignment between two layers. This structure supports the DBE condition when designed carefully.

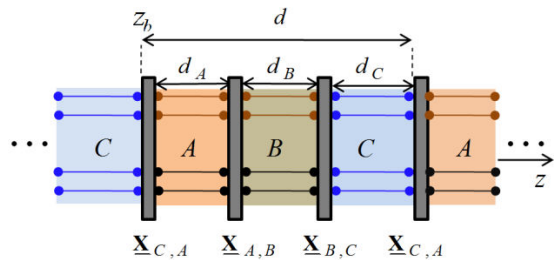

Fig. 3. Schematic of a periodic MTL whose unit cell is made by three segments with two uncoupled TLs. Coupling between TLs occurs at the junction between segments and it is described by a four port network, corresponds to the 4x4 block in the matrix $\underline{\underline{\mathbf{X}}}$ defined in (4). The dispersion diagram of this periodic MTL is capable to support DBE condition. Several periodic waveguides, such as those in Fig. 1, can be modeled with this equivalent circuit.

Here we extend concepts in [31], [32], [34], [35], [51] to a case with two TLs for RF frequencies, as in Fig. 3, that may be related to three cascaded waveguides segments with elliptical tilted cross-sections, as in Fig. 1, with each segment modeled by two equivalent TLs. We will show that this structure supports a DBE. Though there are several other MTL configurations that support the DBE as described in Sec. I, we chose the one in Fig. 3 for its simplicity. Each TL segment in Fig. 3 represents a waveguide segment that supports two *uncoupled* modes. The TL parameters values for the three segments A, B and C are given in Appendix B. In particular we chose the two TLs, in each segment A and B, which are not identical to each other (see TL colors black and brown in Fig. 3), whereas in segment C the two TLs are identical to each other (blue color). Therefore the system has a broken symmetry between the constitutive TLs in layers A and B. Note that this case is equivalent to the one studied in [32] where two dielectric layers are anisotropic, and the third layer is isotropic.

We choose a coupling matrix $\underline{\underline{\mathbf{X}}}_{m,m+1}$ as in (4) to describe the coupling at the junction between two contiguous MTL segments, where $m$ = A, B ,C as

$$\underline{\underline{\mathbf{X}}}_{m,m+1}\left(\delta\varphi_m\right) = \begin{bmatrix} \underline{\underline{\mathbf{Q}}}\left(\delta\varphi_m\right) & \underline{\underline{\mathbf{0}}} & \mathbf{0} & \mathbf{0} \\ \underline{\underline{\mathbf{0}}} & \underline{\underline{\mathbf{Q}}}\left(\delta\varphi_m\right) & \mathbf{0} & \mathbf{0} \\ \mathbf{0} & \mathbf{0} & 1 & 0 \\ \mathbf{0} & \mathbf{0} & 0 & 1 \end{bmatrix},$$   (12)

where $\underline{\underline{\mathbf{Q}}}\left(\delta\varphi_m\right)$ is a 2×2 rotation matrix defined as

$$\underline{\underline{\mathbf{Q}}}\left(\delta\varphi_m\right) = \begin{bmatrix} \cos(\delta\varphi_m) & \sin(\delta\varphi_m) \\ -\sin(\delta\varphi_m) & \cos(\delta\varphi_m) \end{bmatrix},$$   (13)

and $\delta\varphi_m$ is the rotation angle of $\mathbf{V}$ and $\mathbf{I}$ basis, from layer $m$ to layer $m+1$. Note that this matrix rotates both the voltage and current vectors $\mathbf{V}_m$, $\mathbf{I}_m$ as described in Sec. II. Using the notation introduced in (4) we have $\underline{\underline{\mathbf{\Phi}}}^{\mathbf{VV}} = \underline{\underline{\mathbf{\Phi}}}^{\mathbf{II}} = \underline{\underline{\mathbf{Q}}}\left(\delta\varphi_m\right)$ and $\underline{\underline{\mathbf{\Phi}}}^{\mathbf{VI}} = \underline{\underline{\mathbf{\Phi}}}^{\mathbf{IV}} = \underline{\underline{\mathbf{0}}}$. In Appendix A we illustrate how this matrix corresponds to a rotation of electric and magnetic fields pertaining to two distinct modes crossing a junction between two tilted waveguide cross-sections. A geometric description could be as in Fig. 1(a) where the cross sections are elliptical.

In our particular case we assume opposite rotation between C and A segments and A and B segments, i.e., $\delta\varphi_{CA} = -\delta\varphi_{AB}$, whereas we assume no rotation from B to C, i.e., $\delta\varphi_{BC} = 0$ that implies $\underline{\underline{\mathbf{X}}}_{B,C} = \underline{\underline{\mathbf{1}}}$, because this configuration is able to support a DBE as shown in the next subsection. The propagation within each MTL layer A,

B, and C is described using the transfer matrices $\underline{\mathbf{T}}_A$, $\underline{\mathbf{T}}_B$ and $\underline{\mathbf{T}}_C$ respectively, each one obtained from (1) and (3) using individual layer parameters. The transfer matrix of the ABC unit cell in Fig. 3 is then obtained as

$$\underline{\mathbf{T}}(z+d,z) = \underline{\mathbf{T}}_C \ \underline{\mathbf{T}}_B \ \underline{\mathbf{X}}_{A,B} \ \underline{\mathbf{T}}_A \ \underline{\mathbf{X}}_{C,A} \tag{14}$$

A. *Dispersion characteristics of a "cold" MTL periodic structure*

We investigate first the $\omega - k$ dispersion characteristic of an SWS with a DBE, which is plotted in Fig. 4(a), with only real wavenumber; which is the conventional way to represent a dispersion diagram. However, in general the wavenumbers (eigenvalues) of the eigensystem in (9) can be complex (in presence of losses, or when interacting with the beam, or in the electromagnetic band gap of the periodic structure). Therefore, since in this paper we deal with amplification, and to provide a complete picture of the dispersion diagram, we also plot the complex wavenumber with $k = \operatorname{Re}(k) + j\operatorname{Im}(k)$ of modes of the 2-TL periodic structures in Fig. 3, first in the case when it does not couple to the electron beam, i.e., referred here as the "cold" MTL. In the rest of the paper we will plot the complex wavenumber $k = \operatorname{Re}(k) + j\operatorname{Im}(k)$ to identify modes also in presence of the coupling to an electron beam. The 2-TL in Fig. 3 supports two modes in each direction, and using the TL values presented in Appendix B, the dispersion relation in Fig. 4, for the cold structure, is calculated using (5) and (9). For the cold case treated in this subsection, the formulation in (1)-(7) is reduced to the case of a 4-dimensional state vector and the transfer matrix is a 4×4 block. The modal dispersion diagram, comprising the real and imaginary parts of the Bloch wavenumber $k$, shows a high-pass type dispersion that models a variety of coupled-cavity travelling wave tubes [1], [8]. The cut-off frequencies for both modes are designed to be at $\omega d/(2\pi c) = 0.04$, and we observe that modes below such frequency have purely imaginary Bloch wavenumber (a bandgap). Such behaviour is a consequence of the series capacitance in the equivalent distributed circuit in Fig. 2(b). At higher frequencies, we have four solutions of (9) with a purely real Bloch wavenumber, namely $k_1$, $k_2$, $k_3$, $k_4$ in Fig. 4. These modal wavenumbers follow the symmetry conditions due to reciprocity (because this the "cold" case, i.e., without interaction with the electron beam). The symmetry property means that if $k$ is Bloch wavenumber, also $-k$ is a solution of (9), and therefore the four solution satisfy $k_3 = -k_1$, $k_4 = -k_2$, and in Fig. 4 we show that property. At higher frequencies, two modes with wavenumbers $k_1$ and $-k_1$ develop an RBE condition at $\omega d/(2\pi c) = 0.135$, where the Bloch wavenumber reaches the first Brillouin zone edge at $\operatorname{Re}(k_1) = +\pi/d$. The other modes with $k_2$ and $-k_2$ are still propagating at the RBE frequency and higher ones.

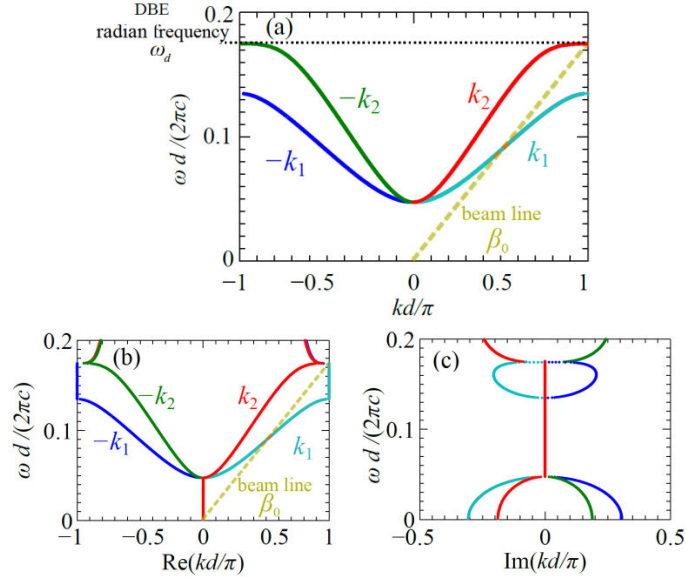

Fig. 4. (a) Dispersion diagram of a periodic structure with a DBE at $\omega = \omega_d$ denoted by the dotted line, showing only the real branches of the Bloch wavenumber. In this example, relative to Fig. 3 and Appendix B, the DBE frequency is at $\omega d / (2\pi c) = 0.175$. Plots of (b) real and (c) imaginary parts of the normalized complex Bloch wavenumber $k$ versus normalized frequency for a "cold" (not coupled to the electron beam) periodic MTL structure supporting two modes at each frequency. The periodic structure is chosen in such a way to develop a regular band edge (RBE) at $\omega d / (2\pi c) = 0.135$ and a degenerate band edge (DBE) at higher frequency ($\omega_d d / (2\pi c) = 0.175$). The yellow dashed line, representing an intrinsic electron beam with velocity $u_0 = \omega / \beta_0$, crosses the elaborated TL dispersion curves at two frequencies where synchronization is qualitatively expected. Note that we designed the system in such a way that the higher-frequency intersection occurs in close proximity with the DBE point.

At higher frequencies, in the region $0.135 < \omega d / (2\pi c) < 0.175$, there exists only two propagating modes with real wavenumbers $k_2$ and $-k_2$, and these modes develop a DBE at $\omega d / (2\pi c) = 0.175$. At higher frequencies than the DBE all modes are in the bandgap, i.e., they all have a wavenumber with an imaginary component. It is important to note that by observing the dispersion diagram one can infer that at the DBE all *four modes have exactly the same Bloch wavenumber*. This means that at the DBE there is only one eigenvalue of (8) with *algebraic multiplicity four*. In mathematical terms, the transfer matrix $\underline{\mathbf{T}}(z+d, z)$ relative to the unit cell of the cold structure (i.e., a 4×4 matrix) becomes similar to a Jordan Block at a DBE. At the DBE the group velocity of the $k_2$ mode is drastically vanishing and this leads to the so called "frozen mode regime". At frequencies $\omega d / (2\pi c) > 0.175$ we have four complex solutions, where each line in Fig. 3(b-c) in that upper frequency region has a multiplicity of two corresponding to complex conjugate wavenumbers.

*Bloch impedance:* Another important quantity of such periodic cold MTL has a Bloch impedance 2×2 matrix defined by $\mathbf{V}(z) = \underline{\underline{\mathbf{Z}}}_B(z)\mathbf{I}(z)$. It is evaluated as detailed in Appendix C, and  based on the general decomposition of modes in a MTL described in [48], [49], [53]. Using similarity transformations, the Bloch impedance is represented as a diagonal $2 \times 2$ matrix such that $\mathbf{V}'(z) = \underline{\underline{\mathbf{Z}}}_{B+}^D(z) \mathbf{I}'(z)$, where the $l^{\text{th}}$ elements of vectors $\mathbf{V}'(z)$, and $\mathbf{I}'(z)$ describe the $n$th eigenvector propagating with Bloch wavenumber $k_n$, and with Bloch impedance $Z_{n+}^D$ being the $i^{\text{th}}$ diagonal element of the diagonal impedance matrix $\underline{\underline{\mathbf{Z}}}_{B+}^D(z) = \mathrm{diag}\left(Z_{1+}^D, Z_{2+}^D\right)$ for a Bloch wave propagating in the positive $z$-direction. For the Bloch waves that propagate in the negative $z$-direction the Bloch impedance is described by a different matrix: $\underline{\underline{\mathbf{Z}}}_{B-}^D(z) = \mathrm{diag}\left(Z_{1-}^D, Z_{2-}^D\right)$. When the Bloch wavenumber $k_i$ is purely real (corresponding to a propagating mode in a lossless structure), the corresponding Bloch impedance $Z_{n\pm}^D$ is in general complex, whereas when $k_n$ is purely imaginary the Bloch impedance is purely imaginary [54]. Moreover, the Bloch impedances $\underline{\underline{\mathbf{Z}}}_{B\pm}^D$ may satisfy properties for symmetric unit cells associated with positive and negative real parts of Bloch wavenumber [55], however in our structures symmetry in the unit cell is broken and therefore the impedance symmetry properties do not hold.

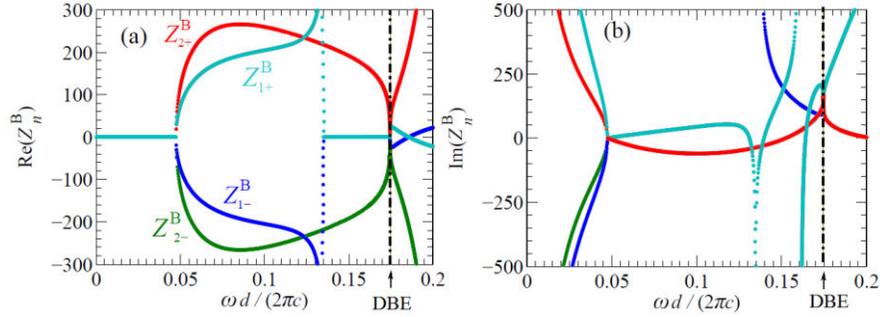

Fig. 5. Plots of (a) real and (b) imaginary parts of the Bloch impedance $Z_{1+}^D$, $Z_{2+}^D$, $Z_{1-}^D$, and $Z_{2-}^D$ versus normalized frequency for the four modes (in Fig. 4) of the "cold" periodic 2-TL structure, evaluated at the beginning of the unit cell shown in Fig. 3. The DBE radian frequency is at $\omega d / (2\pi c) = 0.175$ indicated by the dashed-dotted vertical line, where the real part of the Bloch impedance vanishes.

The Bloch impedance depends on the choice of the $z$-location, and in this case it is evaluated at the beginning of the unit cell shown in Fig. 3 (at the location $z_b$ in Fig. 3). We plot in Fig. 5 the real and imaginary parts of the elements of the diagonal modal Bloch impedance $\underline{\underline{\mathbf{Z}}}_{B\pm}^D$, for the 2-TL case in Fig. 3 with parameters provided in Appendix B. Propagating modes have complex Bloch impedances, and the real part of the impedance vanishes when the mode is in a band gap or at cutoff (see Fig. 5(a)), as shown for example below the cutoff frequency $\omega d / (2\pi c) < 0.05$. Note that a positive (negative) real part of the modal Bloch impedance is related to the power flow across a unit cell in the positive (negative) $z$-direction. It is important to state that exactly at the DBE frequency, there exists only *one eigenvector*, since the transfer matrix is similar to a Jordan block, as discussed in [32], [33]. This degeneracy condition is also manifested in Fig. 5 where the Bloch impedances of the modes (both real and imaginary parts) all converge to a single point at $\omega d / (2\pi c) = 0.175$. It is also observed in Fig. 5 that the real part of the Bloch impedance of the modes vanishes in the vicinity of the DBE, while the imaginary part becomes nonzero.  According to the formulation presented in Appendix C, the definition of diagonal Bloch impedance for MTLs requires presence of independent Bloch modes. However, at the DBE frequency,  since only one Bloch mode is present, a set of independent eigenmodes can be found only by resorting to (non-Bloch) generalized eigenvectors (refer [33], [51] to for detailed discussion on generalized eigenvectors). This only takes place exactly at the DBE and the interaction of the MTL with the electron beam, to be explored next, would significantly alter the impedance at the DBE and at its near vicinity. In the following, we discuss the modification of band diagrams when an electron beam interacts with the periodic MTL.

### B. Dispersion characteristics of a periodic MTL interacting with an electron beam

We now calculate the complex modes for the "hot" MTL structure in Fig. 3 that interacts with the electron beam, as in Fig. 2, by solving the full a 6×6 eigenvalue problem in (9). Therefore we obtain 2$N$+2, with $N$=2,  Bloch modes. They differ from the modes in the cold structure reported above especially in the frequency region where synchronization occurs between the electromagnetic mode(s) and the electron beam. In this example with two TLs the transfer matrix is 6×6 and we obtain six wavenumber solutions. Note that in this case the wavenumber symmetry property does not hold anymore, and in general if $k$ is solution, $-k$ is not. However, this symmetry may be almost satisfied depending on the weakness of the MTL-electron beam interaction.

In Fig. 6 we plot the dispersion diagram with normalized frequency $\omega d / (2\pi c)$  and real and imaginary parts of the normalized Bloch wavenumber $k$ in the case the electron beam has an average (d.c.) velocity $u_0 = 0.34c$ , where $c$ is the speed of light, and the average beam current is  $I_0 = 10 \left[ \text{mA} \right]$ . We report that the dispersion diagram of the coupled system is altered compared to the one in Fig. 4 and we see several complex solutions, hence amplification, due to the synchronism condition. To observe the deformation of the dispersion diagram in this case compared to "cold" dispersion diagram in Fig. 4, we zoom the critical regions.  Amplification is represented by modes with exponential growth, and hence related to the positive imaginary parts of the Bloch wavenumbers. (Not to be confused however with bandgaps.) Note that it has been previously shown in [42] that for the $N$-line uniform system (i.e., homogeneous along $z$) coupled to an electron beam, there exists only one pair (and not more than one, as rigorously proved in [44]) of complex conjugate solutions for the wavenumber. In addition, *the existence of growing and decaying waves is always guaranteed if*

$$\beta_0 \geq \max \left\{ \text{Re}(k_{c,1}), \text{Re}(k_{c,2}), ..., \text{Re}(k_{c,N}) \right\} , \qquad (15)$$

where,  $\pm k_{c,1}, \pm k_{c,2}, ..., \pm k_{c,N}$  are the 2$N$ natural modal wavenumbers of the ″cold″ MTL system [42], discussed in the previous subsection. This also means that in a uniform MTL coupled to an electron beam [42], [44] there exists maximum only one amplification regime via the one and only one exponentially growing mode [44].

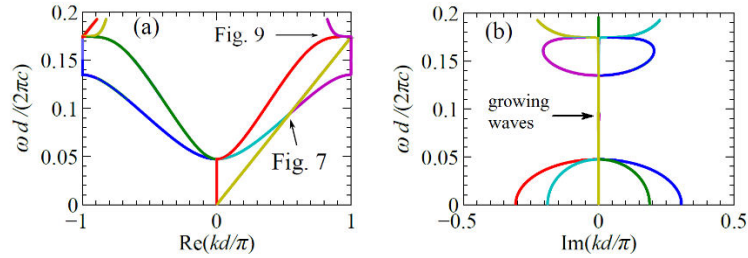

Fig. 6. Plots of (a) real and (b) imaginary parts of the normalized Bloch wavenumber $k$ versus normalized frequency for a system consisting of a periodic structure supporting two modes interacting with an electron beam having an average phase velocity $u_0 = 0.34c$ and $I_0 = 10 \, [\text{mA}]$. The interaction causes deformation of the dispersion diagram, compared to that of the "cold" slow-wave structure in Fig. 4, allowing for growing waves, and zoomed plots provided in Fig. 7 and Fig. 9.

Instead, here, considering periodic structures, we show the possibility of observing more than one exponentially growing solution, *opening the possibility to having more than one mode synchronized to the electron beam co-operating to the amplification process*. In other words, MTL systems with more than one mode synchronized with the electron beam offer different regimes of operation that may even lead to larger amplifications or lager bandwidths than in traditional systems based on single electromagnetic mode interaction. Often the presence of other electromagnetic modes is seen as detrimental, here instead we show a generalized Pierce formulation to account for multiple synchronizations and a regime with more than one mode having exponential growth.

*Single-mode synchronization.* To examine closely the interaction between the two TLs and the electron beam we show a zoomed area at the two interaction regions. First we show the case when the electron beam is synchronized mainly with *only one mode* in the frequency range $0.06 < \omega d /(2\pi c) < 0.12$ in Fig. 6, i.e., with the mode with wavenumber $k_1$, for two different values of the average beam current, $I_0 = 0.1 \, [\text{mA}]$ in Fig. 7(a,b) and $I_0 = 10 \, [\text{mA}]$ in Fig.7(c,d). Because of single mode synchronization we mainly expect a standard TWT way of operation. For brevity, in Fig. 7 we show only the branches with positive real Bloch wavenumber where the possible interaction occurs, while the negative branch of $k$ is not significantly altered in this particular frequency range. Out of the six solutions, four have $\text{Re}(k_n) > 0$ $(n=1,2,3,4)$ and two have $\text{Re}(k_n) < 0$ $(n=5,6)$, though in some frequency region the real part vanishes.

We observe in Fig. 7(a-d) that dispersion curves for the coupled system split in the real and imaginary parts of the Bloch wavenumber as a result of the coupling to the electron beam. This splitting increases by increasing the electron beam current (we maintain constant the beam velocity, therefore we increase the beam charge density), that implies increasing the beam energy. Observe that in this case a mode having $\text{Im}(k) > 0$ indicates a growing wave solution while a mode having $\text{Im}(k) < 0$ denotes a decaying wave solution. (The bandgap would require a more sophisticated description, but here we are interested in amplification regimes). Consequently by comparing Fig. 7(b) with Fig. 7(d) we see that the imaginary part of the Bloch wavenumber increases for increasing beam current. To better understand curves in the dispersion diagram in Fig. 7(c,d) it is convenient to plot modal wavenumbers in the complex plane, varying frequency in the range $0.06 < \omega d /(2\pi c) < 0.12$, for the case when $I_0 = 10 \, [\text{mA}]$. Arrows mean increasing frequency. We only plot the complex half plane with Re($k$)>0 since this is the main region where synchronization with the electron beam occurs. It is clear that for the case shown in Fig. 7(c-d) there are two growing wave solutions (along the positive $z$-direction) at certain frequencies, as also specified in Table I for $\omega d\,/(2\pi c)=0.1$. This interesting feature is revisited later, when discussing the case where the interaction occurs near the DBE of the periodic MTL.

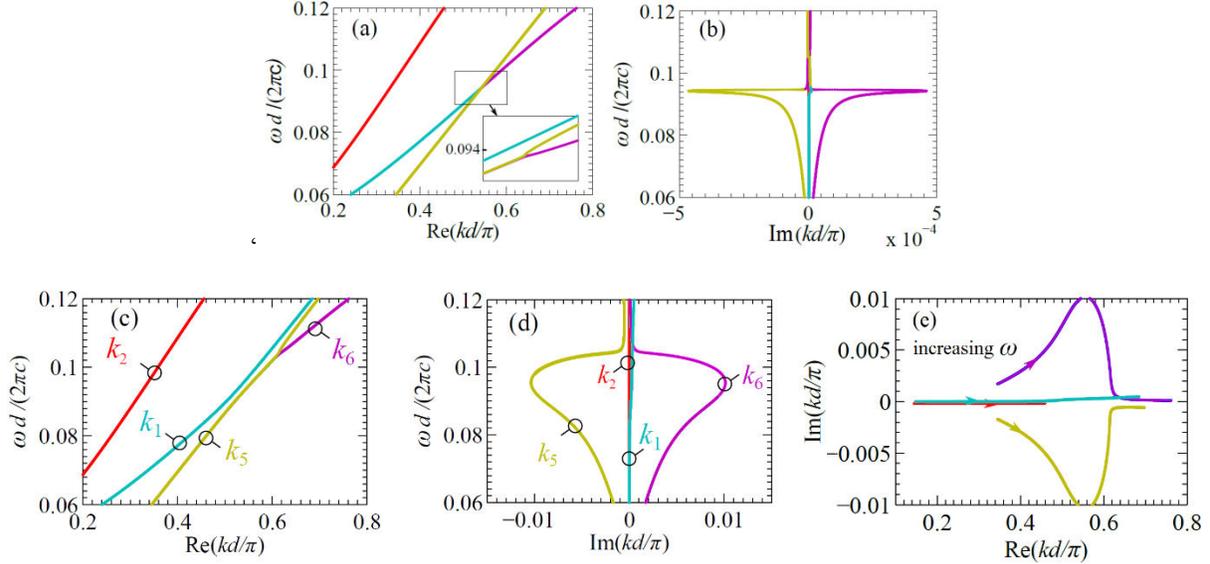

Fig. 7. (a-d) Zoomed plots of real and imaginary parts of the Bloch wavenumber versus normalized frequency for two different values of the average electron beam current: (a-b) $I_0=0.1\,[\mathrm{mA}]$ and (c-d) $I_0=10\,[\mathrm{mA}]$ . We show only the four modal solution with  Re($k$) > 0 since this is the region where the electron beam interacts. In (e) we show the evolution of the wavenumber in complex plane increasing frequency (arrows) for the modes plotted in (c-d) (see arrows).

We consider the state vector $\boldsymbol{\Psi}_1(z)$  corresponding to the Bloch mode with wavenumber $k_1 d\,/\pi=0.565+j2\times10^{-4}$ at  $\omega d\,/(2\pi c)=0.1$. This mode is related to the wavenumber $k_1$ in the cold structure (Fig. 5) and it exhibits now exponential growth along $z$  at  $\omega d\,/(2\pi c)=0.1$ (see light blue curve with Im($k$)>0 in Fig. 7(c-d)). Note that the solution with negative imaginary part is  $k_5 d\,/\pi\approx0.585-j7.5\times10^{-3}$ at  $\omega d\,/(2\pi c)=0.1$, and that now wavenumber symmetry is violated. The other mode that exhibits exponential growth is the mode  $\boldsymbol{\Psi}_6(z)$  with a wavenumber wavenumber $k_6 d\,/\pi=0.585+j7.5\times10^{-3}$ .

**TABLE I. VALUES OF BLOCH WAVENUMBER AT  $\omega d\,/(2\pi c)=0.1$  WITH  $I_0=10\,[\mathrm{mA}]$.**

| mode index $n$ | $n=1$ | $n=2$ | $n=3$ | $n=4$ | $n=5$ | $n=6$ |
|---|---|---|---|---|---|---|

| Re($k_n d / \pi$) | 0.565 | 0.358 | $-0.589$ | $-0.358$ | 0.585 | 0.585 |
|---|---|---|---|---|---|---|
| Im($k_n d / \pi$) | $2 \times 10^{-4}$ | $-7 \times 10^{-7}$ | $-1.2 \times 10^{-7}$ | $-6 \times 10^{-8}$ | $-7.7 \times 10^{-3}$ | $7.5 \times 10^{-3}$ |

Since the MTL-electron beam system is periodic it is important to investigate which Floquet harmonics contribute to the coupling and to the power transportation in the $z$ direction. We calculate the Floquet wave coefficients, of the mode with $k_6$ as shown in Table I, namely, $\boldsymbol{\psi}_6^p = \begin{bmatrix} V_{1,6}^p & V_{2,6}^p & I_{1,6}^p & I_{2,6}^p & V_{b,6}^p & I_{b,6}^p \end{bmatrix}^T$ of the state vector as described in (11). We plot in Fig. 8 the normalized $p$-indexed Floquet weights $v_{l,6}^p = \left| V_{l,6}^p \right| / \left| V_{l,6}^0 \right|$ and $i_{l,6}^p = \left| I_{l,6}^p \right| / \left| I_{l,6}^0 \right|$, with $l$ denoting the transmission line number $l = 1,2$ as well as the charge wave, $n=b$, for the Bloch wave with $k_6$ whose value is found in Table I. The Floquet coefficients are shown in Fig. 8(a,b) for the mode with $n = 6$ that has $k_6 d / \pi = 0.585 + j7.5 \times 10^{-3}$ at $\omega d /(2\pi c) = 0.1$. The normalization values for those Floquet harmonics are the weights corresponding to the $0^{\text{th}}$ harmonic: $V_{1,6}^0 = 1$, $V_{2,6}^0 = 2$, $I_{1,6}^0 = 10$, $I_{2,6}^0 = 20$, $V_{b,6}^0 = 90$, and $I_{b,6}^0 = 0.1$, and all harmonic voltages have units of [V] while current harmonics are in [mA].

We observe that the charge wave is dominated by only the fundamental harmonic (that with $p = 0$) of the Floquet expansion because all the higher harmonics are at least 40 dB weaker, which indicates that the coupling between the electron beam and the TL is occurring only via the fundamental Floquet harmonic of the periodic structure. Because of their weakness, neglecting higher order harmonics in the charge wave would not alter the results.

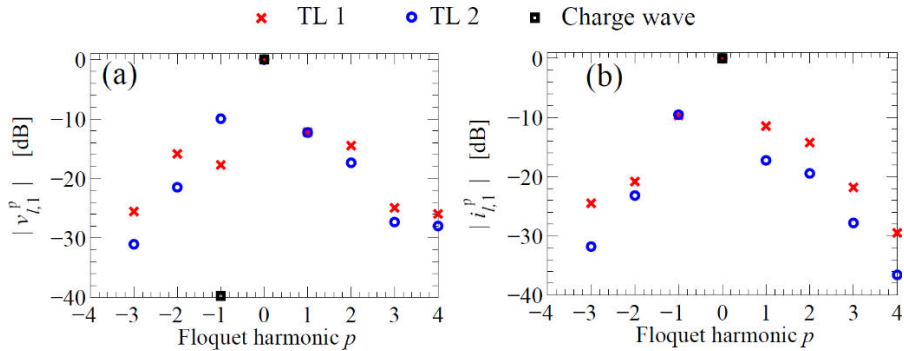

Fig. 8. Plots of normalized Floquet harmonics weights $\boldsymbol{\psi}_6^p = \begin{bmatrix} v_{1,6}^p & v_{2,6}^p & i_{1,6}^p & i_{2,6}^p & v_{b,6}^p & i_{b,6}^p \end{bmatrix}^T$ for the mode with $k_6$, in Table 1. Each mode harmonic component is normalized to its own fundamental harmonic ($p=0$), with $l = 1,2$ denoting the TL number and $l = b$ the charge-wave. The considered state vector is the one for the growing mode at $\omega d /(2\pi c) = 0.1$, with $k_6 d = (0.585 + j0.0073)\pi$.

*C. Super synchronism: synchronization with four TL modes near the DBE.*

Now we investigate the same MTL system as in the previous subsection with the electron beam current $I_0 = 10 \, [\text{mA}]$ considered in the previous case, but we focus on the different scenario pertaining to the frequency region where DBE is manifested in the periodic structure. The MTL system has been designed in such a way the beam line (Fig. 6) intersects the DBE region of the dispersion diagram, a condition we refer to as *super synchronism*. Because two TL Bloch modes have approximately the same wavenumber in DBE region, both modes (the red and purple curves in the "hot" structure dispersion diagram Fig. 6) synchronize with the electron beam. By observing the dispersion of the two electromagnetic modes in the "cold" structure in Fig. 4, this super synchronization involves two modes described by light blue and red curves, denoted by $k_1$ and $k_2$, respectively. However it must be noted that the two curves of these two Bloch modes coalesce at the edge of the Brillouin zone, and therefore the two Bloch modes traveling in opposite direction, with wavenumber $-k_1$ and $-k_2$ (in the cold structure) have Fouquet harmonics that also coalesce at the same point $(k = \pi / d, \omega_d)$. Therefore this super synchronization involves four degenerate Bloch modes, and to be more specific it involves four Floquet harmonics, of their corresponding Bloch modes, that are phase synchronized with the electron beam velocity. Since these four Floquet harmonics have the same wavenumber $k = k_d = \pi / d$, the velocity of the electron beam must be chosen such that

$$u_0 \approx \frac{\omega_d}{k_d} = \frac{\omega_d d}{\pi} , \qquad (16)$$

Due to the complexity of the dispersion diagram in Fig. 6 around the DBE point at $\omega d / (2\pi c) = 0.175$ and because of the vicinity to the second Brillouin zone, in Fig. 9 we show a zoomed plots for both positive and negative branches of the real Bloch wavenumber, along with the imaginary part, in the frequency range $0.16 < \omega d / (2\pi c) < 0.19$. As in the previous case there are six Bloch wavenumber solutions relative to the "hot" structure, three having $\text{Re}(k_n) > 0$ ($n = 1,2,6$) and three having $\text{Re}(k_n) < 0$ ($n = 3,4,6$), and the precise wavenumber values are shown in Table 2 at $\omega d / (2\pi c) = 0.175$. An interesting manifestation of the characteristic modes of such periodic structure is that, in contrast to the uniform MTL interacting system [37], we find more than one propagating mode that grow exponentially, that are able to contribute to amplification and to the energy flow along the +z-direction.

**TABLE II. DBE CASE: VALUES OF BLOCH WAVENUMBER AT** $\omega d / (2\pi c) = 0.175$ **WITH** $I_0 = 10 \, [\text{mA}]$.

| mode index $n$ | $n = 1$ | $n = 2$ | $n = 3$ | $n = 4$ | $n = 5$ | $n = 6$ |
|---|---|---|---|---|---|---|
| $\text{Re}(k_n d / \pi)$ | 0.957 | 0.9216 | $-0.962$ | $-0.917$ | $-0.962$ | 0.9672 |
| $\text{Im}(k_n d / \pi)$ | 0.0627 | $-0.0051$ | 0.0737 | 0.0024 | $-0.0671$ | $-0.0667$ |

We first notice in Fig. 9(b,c) an exponentially growing ( $k_1$ , light blue colored curve). This mode has $\mathrm{Re}(k) > 0$ and a large positive $\mathrm{Im}(k)$. Though not shown here, by increasing the beam energy, the imaginary part further increases. We also observe in Fig. 9(a,c) another exponentially growing mode (i.e., with $\mathrm{Im}(k) > 0$ ) around the DBE frequency, in the region which has $\mathrm{Re}(k) < 0$ ( $k_4$ , green curve). This mode however has relatively smaller $\mathrm{Im}(k)$ than the other exponentially growing mode. The frequency evolution of modal wavenumbers is better understood in the complex $k$-plane representation in Fig. 10. There arrows indicate increasing frequency, and the frequency range is the same as that considered in Fig.9. All modes, though small in some cases, have a non-zero imaginary part of $k$. The exponentially growing mode denoted by $k_1$, with a light blue curve in Fig. 9(b,c) has a real part that increases with frequency, but does not reach the Brillouin zone edge (where $\mathrm{Re}(kd) = \pi$ ) and it rather remains positive with $\mathrm{Re}(kd) < \pi$ also for higher frequencies. These results show that in this frequency range the electron beam has a significant impact on the modal dispersion of the "cold" structure and the degeneracy condition is no longer strictly manifested because the four eigenvalues (i.e., the four wavenumbers) are not exactly equal to each other, as it happens instead for the "cold" structure.

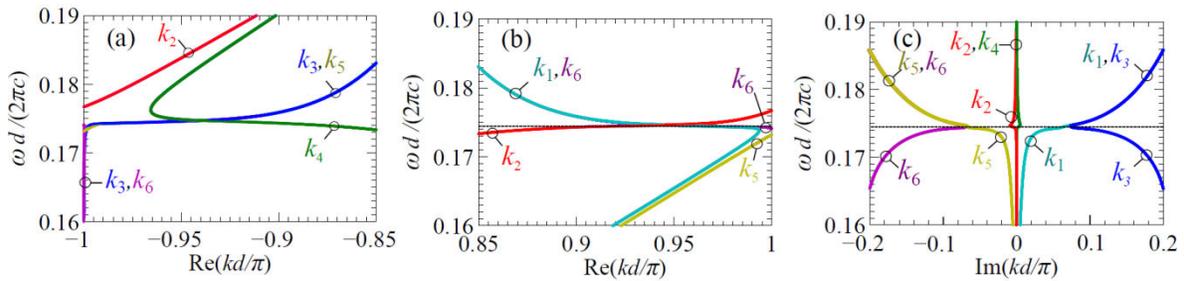

Fig. 9. Super synchronism condition, four modes with DBE interacting with the electron beam. Zoomed plots of (a-b) real, positive and negative, and (c) imaginary parts of the Bloch wavenumber $k$ in Fig. 6, for the frequency range $0.16 < \omega d / (2\pi c) < 0.19$, assuming a beam current of $I_0 = 10 \, [\mathrm{mA}]$. Note that the real part of the Bloch wavenumber is separated in two plots, both close to the edge of the Brillouin zone. Furthermore, near the edge of the Brillouin zone we have four TL modes that close to degeneracy, and all interacting with the electron beam.

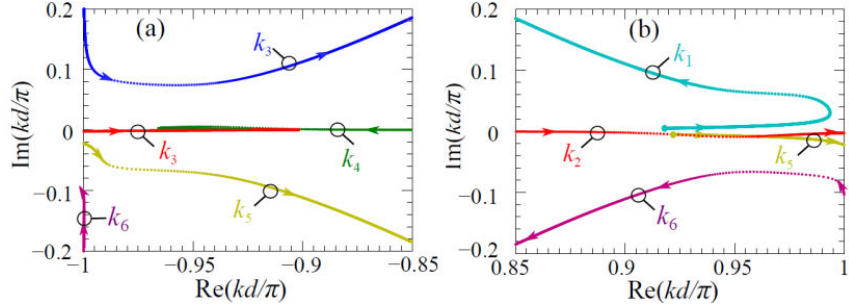

Fig. 10. Super synchronism condition, four modes with DBE interacting with the electron beam. Plot of the wavenumber evolution in the complex plane with increasing frequency (indicated by arrows), in the frequency range considered in Fig. 9. The starting frequency is indicated by a dot. Note that the complex plane is separated into two: the positive and negative real branches of $k$, corresponding to Figs. 9(a) and (b), respectively.

Since synchronization between the electron beam and the MTL occurs in the DBE region, with four modes, a strong interaction is expected and it is insightful to observe the deformation of the dispersion curves as a function of the electron beam intensity, shown in Fig. 11. For clarity, we show what happens at the edge of the Brillouin zone by plotting the real part of $k$ just before and after the band edge, in the normalized frequency range $0.9\pi < kd < 1.1\pi$, to show how modal curves split in that region. Notice how the DBE in the "cold" MTL (dotted curves) is perturbed by beam perturbation. For electron beam d.c. current $I_0 = 10 \, [\text{mA}]$ we see that the DBE condition (solid curves) is deteriorated, compared to the "cold" MTL, and the modes no longer satisfy the relation $\Delta\omega \sim (\Delta k)^4$ that is typical of DBE, however for frequencies away from the DBE (not shown in Fig. 11) the four modes asymptotically approach those of the cold structure. Increasing the beam average current $I_0$ results in higher deformation of the dispersion diagram, and for a current of $I_0 = 1 \, [\text{A}]$ the DBE typical dispersion (solid curves) cannot be recognized anymore, and the advantages relative to it may be lost. This stress the importance of maintaining small perturbation of the DBE (in terms of gain) in order to fully utilize the possible advantages of the super synchronous regime. The idea of using small perturbation to the DBE to achieve highest possible gain is evident, which was expressed in [36] in the context of finite active structures operated at the DBE that are able to produce a giant gain only when the DBE is not largely perturbed. Another important characterization of the interactive system, which is omitted from this paper, is the detailed analysis of the energy transfer between the electron beam and modes and their Floquet harmonics. A Lagrangian treatment for the energy exchange in the system can be developed [44], or equivalently by employing the energy conservation and Poynting theorem as done in [42] for homogenous MTL interacting with the electron beam.

For this peculiar synchronization regime involving four degenerate TL modes we report in Fig. 12 the weight of the magnitude of Floquet harmonics for one of the modes that is exponentially growing at $\omega d \, / (2\pi c) = 0.1745$, corresponding to the light blue curve in Fig. 9, with $k_1$ given in Table II. Voltage and current Floquet harmonics $V_{l,1}^p$, with $l=1,2,b$, are normalized to their own strongest $p^{\text{th}}$ harmonics, that are reported here:

$\left|V_{1,1}^p\right|_{\max} = \left|V_{1,1}^{-1}\right| = 0.0069\,[\text{V}]$, $\left|V_{2,1}^p\right|_{\max} = \left|V_{2,1}^0\right|_{\max} = 0.0047\,[\text{V}]$, $\left|I_{1,1}^p\right|_{\max} = \left|I_{1,1}^{-1}\right|_{\max} = 0.046\,[\text{mA}]$, $\left|I_{1,1}^p\right|_{\max} = \left|I_{1,1}^0\right|_{\max} = 0.052\,[\text{mA}]$.

$\left|V_{b,1}^p\right|_{\max} = \left|V_{b,1}^0\right|_{\max} = 0.1\,[\text{V}]$, and $\left|I_{b,1}^p\right|_{\max} = \left|I_{b,1}^0\right|_{\max} = 2.25\times10^{-4}\,[\text{mA}]$. The charge wave is dominated by the fundamental

Floquet harmonic ($p$=0) since higher order harmonics have weights lower than 40 dB. Voltage and current in each of the two TLs have highest contribution from both the $p = -1$ and the $p = 0$ harmonics, whereas higher order harmonics' weight slightly decreasing with harmonic $p$-index. The reason why the $p = -1$ harmonic is involved is because the DBE condition occurs at the boundary of the Brillouin zone, involving modes that belong to different

Brillouin zones. It is important to note that at the DBE in the "cold" periodic structure, the Bloch mode (voltage and current) is mainly represented by two Floquet harmonics with equal weights.

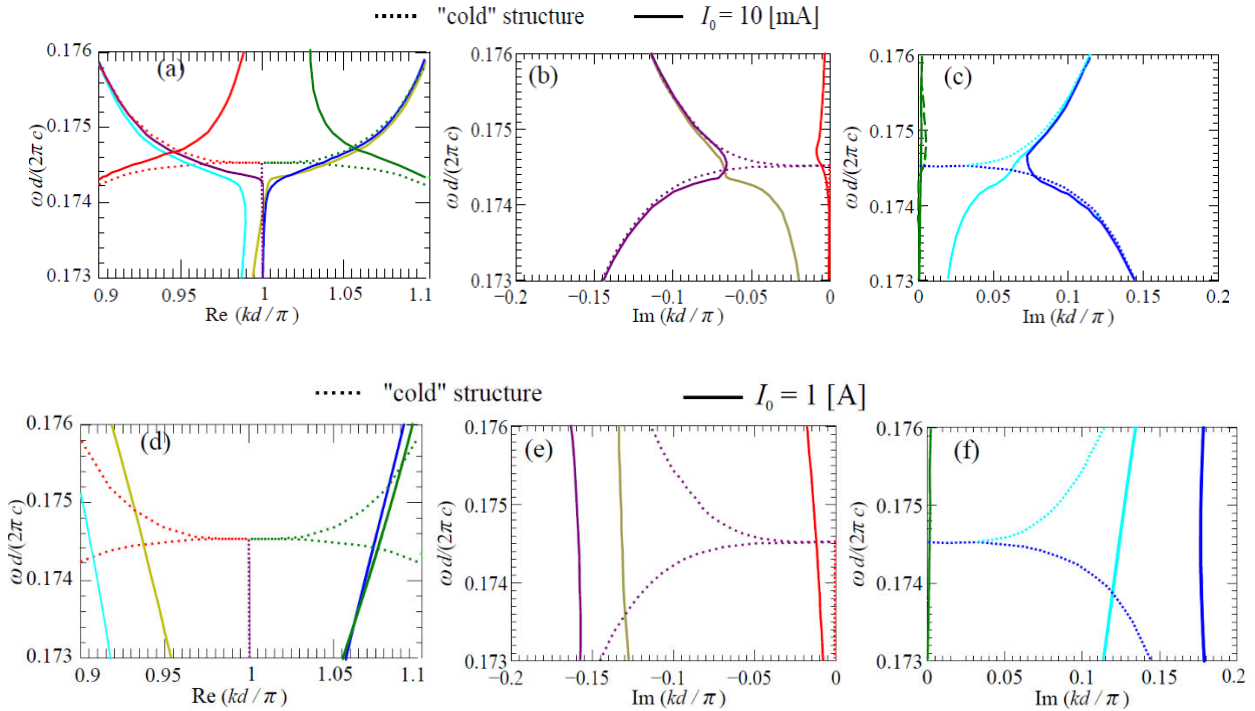

Fig.  11. Super synchronism condition, four modes with DBE interacting with the electron beam. Plots of real (a,d)  and imaginary (b-c) and (e,f) parts of the Bloch wavenumbers $k$ versus frequency for two different electron beam average currents

(solid lines) as well as for the "cold" MTL structure in Fig. 3 that exhibits a DBE at $\omega d / (2\pi c) = 0.175$ (dotted lines). Note  the real branches are plotted here at the interface between the fundamental Brillouin zone and the next to show the continuity of mode curves at the band edge, close to the DBE. We show dispersion curves with both Im($k$)<0 and Im($k$)>0 in (b,e), and (c,f), respectively.

Therefore, when coupled to the e-beam, the interactive-system modes, operating very close to the DBE condition,

are also well-represented by two main Floquet harmonics. Yet, the 0th Floquet harmonic is the one interacting with the charge wave harmonic, as seen in Fig. 12. This also suggest  the possibility of utilizing the DBE for designing backward wave oscillators [25], [56] when careful dispersion engineering is carried out via synchronism close the DBE with different Floquet harmonics. A rigorous study of the coupling mechanism of the different spatial harmonics and their contribution to amplification and oscillation will be carried out in future investigations.

## IV. MODES IN PERIODIC MTL SYSTEMS: DBE CASE USING PERIODIC TL SECTIONS WITH MODE MIXING

In his section we study the interaction of an electron beam with another example of a periodic structure whose dispersion diagram can develop a DBE. An illustrative example based on mode mixing is shown in Fig. 13, where the unit cell of length $d$ is composed of two TL segments, with two TLs in each segment. In this case TLs' coupling does not occur at TL junctions, i.e., $\underline{\mathbf{X}}_{AB} = \underline{\mathbf{X}}_{BA} = \underline{\mathbf{1}}$. However, we now assume that within the $B$ segments, TLs have intrinsic coupling by inductive and/or capacitive means, contrary to the structure in Fig. 3 where there is no

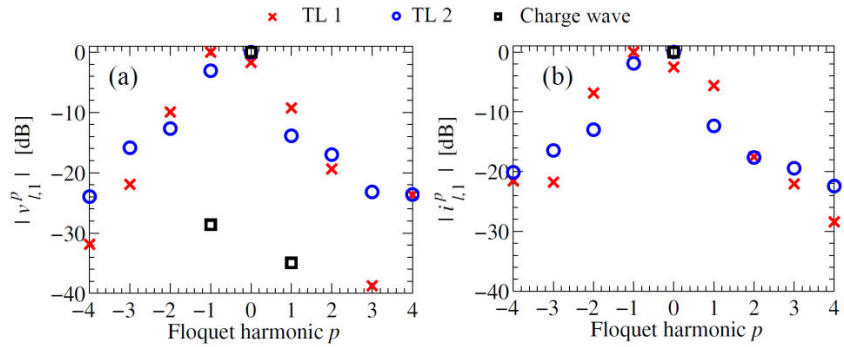

Fig. 12. (a-b) Plots of normalized Floquet harmonics weights $\boldsymbol{\psi}_1^p = \begin{bmatrix} v_{1,1}^p & v_{2,1}^p & i_{1,1}^p & i_{2,1}^p & v_{b,1}^p & i_{b,1}^p \end{bmatrix}^T$ for the mode with $k_1 d = (0.957 + j0.0627)\pi$ in Table II, which is a growing mode near the DBE, at $\omega d / (2\pi c) = 0.175$. Voltages and currents are normalized to their strongest harmonic (either the $p = 0$ or the $p = -1$ harmonic). The strongest harmonic of the charge wave is always the one with $p = 0$ and the others can always be neglected.

Coupling inside MTL segments. This case has some analogy to the microstrip TLs in [38], [39], [57] where a segment with two uncoupled and parallel microstrips is directly connected to another segment with two coupled (closely spaced) and parallel microstrips. The same concept can be realized with two coupled rectangular or circular waveguides. Note that in the distributed circuit parameters of the MTL in Fig. 2, the electron beam is seen as a current injection  in the transmission lines in accord to [4,5] and [38], and it is assumed implicitly from this model that the electron beam does not inject current in the coupling capacitances or inductances. We can readily obtain the transfer matrix of the unit cell as

$$\underline{\mathbf{T}}(z + d, z) = \underline{\mathbf{T}}_B \underline{\mathbf{T}}_A = e^{-\underline{\mathbf{M}}_B d_B} e^{-\underline{\mathbf{M}}_A d_A} \ . \tag{17}$$

In this example, we design a particular MTL system with segment $A$ composed of uncoupled 2 TLs while segment B has intrinsic inductive and capacitive coupling. It is also designed such that the dispersion relation of the "cold" MTL develops a DBE at $\omega d / (2\pi c) = 0.156$  using the TL parameters in Appendix B.

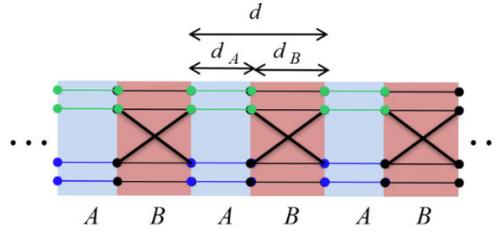

Fig. 13. Periodic structure composed of cascading 2 segments of MTL without the coupling interface **X**. In this particular structure, the unit cell is composed of two segments of MTL, one is uncoupled and the other is inductively and/or capacitively coupled. The dispersion diagram of the structure can develop a DBE.

We plot in Fig. 14 the dispersion diagram of the "hot" structure composed of two segments of MTLs as described by the transfer matrix in (14) interacting with an electron beam, whose electrons have an average velocity $u_0 = 0.3\mathrm{l}c$, so satisfying the *super synchronism* condition (16), and the beam current is $I_0 = 15\,[\mathrm{mA}]$. The zoomed plots in Fig. 15 show the dispersion diagram near the DBE frequency. The same features observed in Fig. 7 are manifested also in this configuration, where electron beam is synchronous to four MTL Bloch modes near the DBE frequency (to be specific, it is synchronous to four Floquet harmonics with wavenumber close to $k = \pi/d$, according to (16)). The dispersion diagram of Im($k$) in Fig. 15(c) shows multiple growing modes near the DBE frequency, thanks to the multi-synchronism involving more than one TL mode.

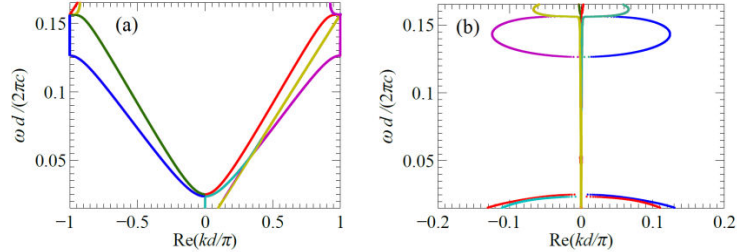

Fig. 14. Plots of (a) real and (b) imaginary parts of the complex Bloch wavenumber $k$ for the periodic structure in Fig. 13 interacting with an electron beam with average velocity $u_0 = 0.3\mathrm{l}c$ and an average current $I_0 = 15\,[\mathrm{mA}]$. The "cold" MTL structure has a DBE at $\omega d/(2\pi c) = 0.156$, and a zoomed plot around that frequency for the *interactive* system in shown in Fig. 15.

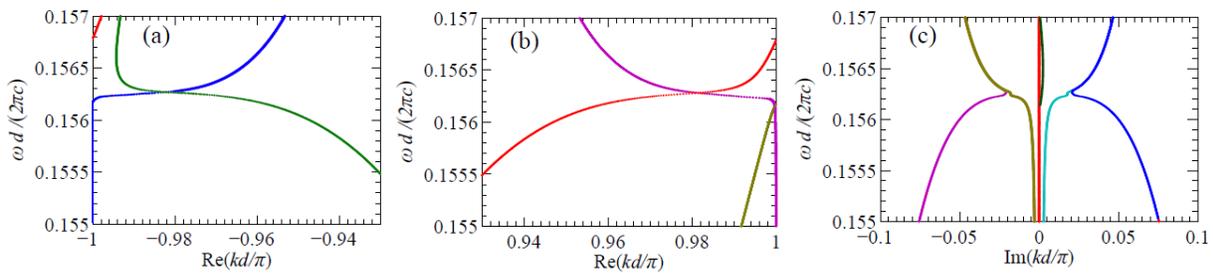

Fig. 15. Super synchronism condition, four modes with DBE interacting with the electron beam. Zoomed plots of (a-b) real and (c) imaginary parts of Bloch wavenumber $k$ for the case in Fig. 14, around the DBE frequency $\omega d/(2\pi c)=0.156$. More than one mode is synchronized with the electron beam generating multiple exponentially growing regimes.

## V. COMPARISON TO A CONVENTIONAL TYPE TWT

Finally we outline here some potential advantages of using periodic structures with a DBE over conventional Pierce type (single mode synchronization) TWT amplifiers in terms of gain. We assume here a periodic TL of finite length $L=Nd$, where $N$ is the number of unit cells and $d$ is their physical length, we calculate the power gain in the interactive structure, and compare it to Pierce type TWT [5] keeping the same average MTL parameters, see Appendix B. We first assume the following boundary conditions: A generator described by the vector $\mathbf{v}_g = \begin{bmatrix} v_{g1} & v_{g2} \end{bmatrix}^T$ is located at $z=0$ as shown in Fig. 16. At a distance $L=Nd$ the MTL is terminated by a 2×2 impedance matrix $\underline{\underline{\mathbf{Z}}}_L$. We further assume that the electron beam enters the interaction area at $z=0$ without any pre-modulation, i.e., we assume $I_b(0)=0$ and $V_b(0)=0$. Then we define the power gain $G$ of the DBE TWT as the ratio between the time-averaged power delivered to the load impedance to the power supplied by the voltage generators,

$$G = \frac{\frac{1}{2}\mathrm{Re}\left(\mathbf{V}^T(L)\mathbf{I}^*(L)\right)}{\frac{1}{2}\mathrm{Re}\left(\mathbf{v}_g^T\mathbf{I}^*(0)\right)} = \frac{\mathrm{Re}\left(\mathbf{I}^T(L)\underline{\underline{\mathbf{Z}}}_L^T\mathbf{I}^*(L)\right)}{\mathrm{Re}\left(\mathbf{v}_g^T\mathbf{I}^*(0)\right)}. \qquad (18)$$

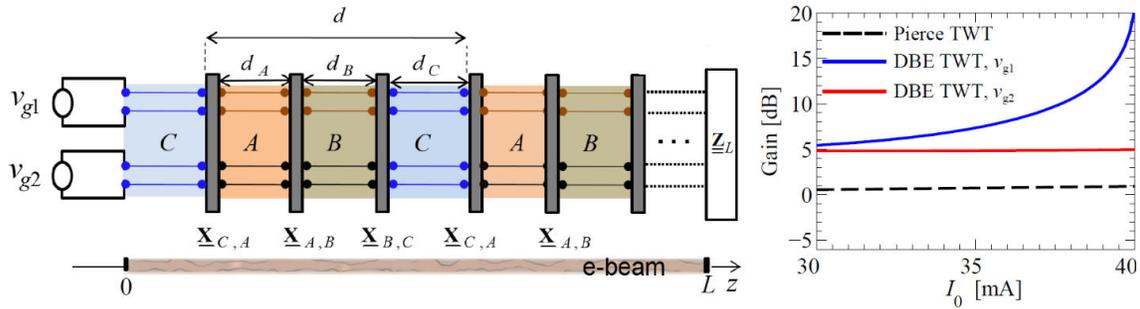

Fig. 16. Comparison between the gain in two finite length TWTs: MTL periodic structure in Fig. 3 made of $N=10$ unit cells, operating with super synchronism at the DBE frequency (solid) for two different generator excitations, and the Pierce type conventional TWT (dashed), assuming the length of the interaction area to be equal to the length $L$ of the DBE TWT.

We compute the gain (18) for a DBE TWT with $N=10$ unit cells whose details are given in in Fig. 3, with each TL terminated by a 50 [$\Omega$] impedance, in other words $\underline{\underline{\mathbf{Z}}}_L$ is diagonal with elements equal to 50 [$\Omega$]. For comparison, we compute the gain also for a conventional Pierce type TWT for which we assume a single uniform TL (Appendix B) that has a characteristic impedance of 50 [$\Omega$] with the same electrical length (the electrical length here is defined as the ratio of the physical length $L$ to the wavelength of the charge wave $2\pi/\beta_0$ at the DBE frequency, which is written as $L/(2\pi/\beta_0)=Nd\beta_0/(2\pi)$ as done in [2], [58]), interacting with the electron beam with the same parameters as for the DBE case. In Fig. 16 we plot the gain in DBE TWT for the two cases with $\mathbf{v}_g = \begin{bmatrix} v_{g1} & 0 \end{bmatrix}^T$ and $\mathbf{v}_g = \begin{bmatrix} 0 & v_{g2} \end{bmatrix}^T$ at the DBE frequency $\omega_d d / (2\pi c) = 0.175$, and also for the single uniform Pierce-type TL TWT. The qualitative gain enhancement is up to 10 times shown in Fig. 16 for one excitation condition $\mathbf{v}_g = \begin{bmatrix} v_{g1} & 0 \end{bmatrix}^T$ compared to Pierce type TWT in a narrow beam current range from 30 to 40 mA. This is caused by the reduced group velocity at the DBE and the fact that more than one mode of the coupled system is exponentially growing. This postulates that with the same beam energy that is relatively small, a DBE TWT can provide ~10 times higher gain that a conventional single TL TWT. It is important to note that for larger beam current, i.e., $I_0 > 40 \, [\text{mA}]$, we observe that the structure may enter a self-excitation (oscillation) regime, owing to the very large quality $Q$-factor of the structure that potentially allows a significant feedback mechanism that leads to oscillations [36]. (The very large $Q$-factor is generated by the fact that the two 50 $[\Omega]$ impedances do not form a Bloch impedance and hence the finite structure is mismatched.) The oscillation starting current in this regime is in the order of 500 mA and it is found by observing the complex pole locations of the transfer function of the finite length DBE structure [56]. These qualitative properties suggest potential use of such structure in oscillators with low starting current and very high efficiency and this will thoroughly investigated in a future study.

## VI. CONCLUSION AND DISCUSSION

We have developed a theoretical framework for the analysis of multimodal slow wave structures, modeled as periodic multi transmission lines that interact with an electron beam following the linearized pierce theory. Two MTL configurations have been considered that are able to model periodic waveguide structures that support multiple modes and can exhibit a special degeneracy condition of the fourth order at a certain frequency. We have calculated dispersion diagram of the Bloch wavenumber as well as the Bloch impedance for some selected case where four electromagnetic modes exhibit a special degeneracy condition at the band edge, called the "degenerate band edge" or DBE. Under this condition, we have explored a super synchronism regime where the electron beam is synchronized with the four degenerate Bloch modes (i.e., with a spatial harmonic of each Bloch mode). We have observed amplification regimes where multiple modes grow exponentially. This is in contrast with the original Pierce model for a single TL where only one mode is allowed to grow exponentially. It is also in contrast to the uniform MTL system interacting with an electron beam where rigorously only one mode is allowed to have exponential growth. A Floquet wave investigation lead to the understanding of which spatial harmonics are strongly excited and which one is responsible for the energy exchange between the slow wave structure and the electron beam. This analysis shows a potential advantage in terms of gain when the super synchronism condition of four modes in the cold structure is achieved and open the way to future investigations related to multimode synchronism.

Travelling wave tubes operated near the DBE can be a candidate for gigantic gain boost, since the degeneracy condition allows for super synchronous operation and hence for a strong beam - RF interaction. A slow wave structure based on the DBE condition can be utilized also as high $Q$ factor resonator with enhanced power conversion efficiency when used as an oscillator. High efficiency amplifier and oscillators based on the DBE can be readily optimized owing to to the use of feeble electron beam energy so as not to significantly perturb the fine characteristics of the DBE condition still resulting in large gain.

## ACKNOWLEDGMENT

This research was supported by AFOSR MURI Grant FA9550-12-1-0489 administered through the University of New Mexico. The authors would like to thank A. Figotin and G. Reyes at the University of California, Irvine for fruitful and intense discussions. F.C acknowledges stimulating discussions with Giovanni Soffici, University of Florence, Italy, years ago that made them very excited about the physics of travelling wave tubes.

## APPENDIX A.
## FIELD REPRESENTATION AND MULTIPLE TRANSMISSION LINES

We consider a medium where two field are able to propagate along the $z$-direction, along with the charge wave accompanied by the electron beam. This procedure can also describe more than two modes, up to $N$ (see [59]) however for the sake of brevity we only focus on two modes that pertain to the structure in Fig. 3. The medium is bounded in the transverse-to-$z$ direction, and it is made of a cascade of stacks of waveguides regions (or segments) as the example shown in Fig. 1. In each segment, let $\mathbf{E}_{t,1}(\mathbf{r})$ and $\mathbf{E}_{t,2}(\mathbf{r})$ be the transverse components of the electric field relative to two modes supported by the waveguide segments. Based on the separation of variable in each segment, the transverse electric field is represented as

$$\mathbf{E}_{t,1}(\mathbf{r}) = \mathbf{e}_1(\boldsymbol{\rho})V_1(z), \text{ and } \mathbf{E}_{t,2}(\mathbf{r}) = \mathbf{e}_2(\boldsymbol{\rho})V_2(z), \tag{B1}$$

where $\mathbf{r} = \boldsymbol{\rho} + z\hat{\mathbf{z}}$, and $\boldsymbol{\rho} = x\hat{\mathbf{x}} + y\hat{\mathbf{y}}$. Analogously we have for the magnetic field

$$\mathbf{H}_{t,1}(\mathbf{r}) = \mathbf{h}_1(\boldsymbol{\rho})I_1(z), \text{ and } \mathbf{H}_{t,2}(\mathbf{r}) = \mathbf{h}_2(\boldsymbol{\rho})I_2(z), \tag{B2}$$

where $\mathbf{e}_n(\boldsymbol{\rho})$ and $\mathbf{h}_n(\boldsymbol{\rho})$ are the electric and magnetic modal eigen-functions, respectively and $V_n$ and $I_n$ are the amplitudes of those fields that describe the evolution of electromagnetic waves along the z-direction. We can assume that $\mathbf{e}_1(\boldsymbol{\rho})$ and $\mathbf{e}_2(\boldsymbol{\rho})$ are orthonormal eigen-functions (that means they are orthogonal with an inner product defined in the cross section with a norm equals to one), and consequently $\mathbf{h}_1(\boldsymbol{\rho})$ and $\mathbf{h}_2(\boldsymbol{\rho})$ . Now we may write the total transverse electric field in the two contiguous segments A and B (in Fig. 3) as

$$\mathbf{E}_{t,A} = \mathbf{e}_{A,1}(\boldsymbol{\rho})V_{A,1}(z) + \mathbf{e}_{A,2}(\boldsymbol{\rho})V_{A,2}(z),$$
$$\mathbf{E}_{t,B} = \mathbf{e}_{B,1}(\boldsymbol{\rho})V_{B,1}(z) + \mathbf{e}_{B,2}(\boldsymbol{\rho})V_{B,2}(z), \tag{B3}$$

and similarly for the transverse magnetic fields

$$\mathbf{H}_{t,A} = \mathbf{h}_{A,1}(\boldsymbol{\rho})I_{A,1}(z) + \mathbf{h}_{A,2}(\boldsymbol{\rho})I_{A,2}(z),$$
$$\mathbf{H}_{t,B} = \mathbf{h}_{B,1}(\boldsymbol{\rho})I_{B,1}(z) + \mathbf{h}_{B,2}(\boldsymbol{\rho})I_{B,2}(z). \tag{B4}$$

The boundary conditions between those two contiguous segments dictates the continuity of the transverse fields, which reads for the electric fields

$$\mathbf{e}_{A,1}(\boldsymbol{\rho})V_{A,1}(z_{AB}) + \mathbf{e}_{A,2}(\boldsymbol{\rho})V_{A,2}(z_{AB}) = \mathbf{e}_{B,1}(\boldsymbol{\rho})V_{B,1}(z_{AB}) + \mathbf{e}_{B,2}(\boldsymbol{\rho})V_{B,2}(z_{AB}). \tag{B5}$$

We take the scalar product of each side with $\mathbf{e}_{A,1}(\boldsymbol{\rho})$ and $\mathbf{e}_{A,2}(\boldsymbol{\rho})$, that are assumed orthonormal, leading to

$$V_{A,1}(z_{AB}) = \left(\int_S \mathbf{e}_{B,1} \cdot \mathbf{e}^*_{A,1} ds\right) V_{B,1}(z_{AB}) + \left(\int_S \mathbf{e}_{B,2} \cdot \mathbf{e}^*_{A,1} ds\right) V_{B,2}(z_{AB}), \tag{B6}$$

$$V_{A,2}(z_{AB}) = \left(\int_S \mathbf{e}_{B,1} \cdot \mathbf{e}^*_{A,2} ds\right) V_{B,1}(z_{AB}) + \left(\int_S \mathbf{e}_{B,2} \cdot \mathbf{e}^*_{A,2} ds\right) V_{B,2}(z_{AB}). \tag{B7}$$

Analogously for the magnetic field equations

$$I_{A,1}(z_{AB}) = \left(\int_S \mathbf{h}_{B,1} \cdot \mathbf{h}^*_{A,1} ds\right) I_{B,1}(z_{AB}) + \left(\int_S \mathbf{h}_{B,2} \cdot \mathbf{h}^*_{A,1} ds\right) I_{B,2}(z_{AB}), \tag{B8}$$

$$I_{A,2}(z_{AB}) = \left(\int_S \mathbf{h}_{B,1} \cdot \mathbf{h}^*_{A,2} ds\right) I_{B,1}(z_{AB}) + \left(\int_S \mathbf{h}_{B,2} \cdot \mathbf{h}^*_{A,2} ds\right) I_{B,2}(z_{AB}). \tag{B9}$$

The charge wave is assumed to be invariant across that interface, such that the equivalent beam voltage and current $I_b$ and $V_b$ remain unchanged.  Using the state vector defined in each segment as

$$\boldsymbol{\Psi}_A(z_{AB}) = \begin{bmatrix} V_{A,1} & V_{A,2} & I_{A,1} & V_{A,2} & V_b & I_b \end{bmatrix}^T, \quad \boldsymbol{\Psi}_B(z_{AB}) = \begin{bmatrix} V_{B,1} & V_{B,2} & I_{B,1} & V_{AB,2} & V_b & I_b \end{bmatrix}^T. \tag{B10}$$

The continuity equation of fields at the interface point $z_{AB}$ can be cast in the following matrix form

$$\boldsymbol{\Psi}_A(z_{AB}) = \underline{\mathbf{X}}_{AB} \boldsymbol{\Psi}_B(z_{AB}) \tag{B11}$$

where we use (B6) through (B9) to write

$$\underline{\mathbf{X}}_{AB} = \begin{bmatrix} \langle \mathbf{e}_{B,1}, \mathbf{e}_{A,1} \rangle & \langle \mathbf{e}_{B,2}, \mathbf{e}_{A,1} \rangle & 0 & 0 & 0 & 0 \\ \langle \mathbf{e}_{B,1}, \mathbf{e}_{A,2} \rangle & \langle \mathbf{e}_{B,2}, \mathbf{e}_{A,2} \rangle & 0 & 0 & 0 & 0 \\ 0 & 0 & \langle \mathbf{h}_{B,2}, \mathbf{h}_{B,2} \rangle & \langle \mathbf{h}_{B,1}, \mathbf{h}_{A,2} \rangle & 0 & 0 \\ 0 & 0 & \langle \mathbf{h}_{B,2}, \mathbf{h}_{A,1} \rangle & \langle \mathbf{h}_{B,1}, \mathbf{h}_{A,1} \rangle & 0 & 0 \\ 0 & 0 & 0 & 0 & 1 & 0 \\ 0 & 0 & 0 & 0 & 0 & 1 \end{bmatrix}, \tag{B12}$$

using the notation of the inner product between arbitrary vectors in the complex normed space by $\langle \mathbf{A}, \mathbf{B} \rangle = \int_S \mathbf{A} \cdot \mathbf{B}^* ds$, thereby finding the bases rotation, or coupling matrix $\underline{\mathbf{X}}_{AB}$ in (B12). As such, by evaluating the eigenmodes projections we find that in general

$$\underline{\mathbf{X}}_{m,m+1} = \begin{bmatrix} \underline{\boldsymbol{\Phi}}_m^{\mathbf{VV}} & \underline{\boldsymbol{\Phi}}_m^{\mathbf{VI}} & \mathbf{0} & \mathbf{0} \\ \underline{\boldsymbol{\Phi}}_m^{\mathbf{IV}} & \underline{\boldsymbol{\Phi}}_m^{\mathbf{II}} & \mathbf{0} & \mathbf{0} \\ \mathbf{0} & \mathbf{0} & 1 & 0 \\ \mathbf{0} & \mathbf{0} & 0 & 1 \end{bmatrix}. \tag{B13}$$

In general, the $\underline{\underline{\Phi}}$ block matrices can constitute up to $N \times N$ mode accounting for multiple propagating and evanescent modes (in the waveguide segments) that may be excited at the discontinuities. However for simplicity, we adopt only two modes in all waveguide segments through this paper. A simple case for interface matrix is shown in (12) as a rotation matrix.

## APPENDIX B.

## MTL PARAMETERS USED IN THE NUMERICAL SIMULATIONS

In all cases, we assume $\mathbf{s} = \mathbf{a}$ with $s_n = a_n = 1$, for $n = 1,2$ and no losses nor space charge effects, i.e., $\underline{\underline{\mathbf{R}}}_m = 0$ and the plasma frequency $\omega_p = 0$ (no space charge effect) unless specified. The plots in Sec. III are generated using the following parameters for a unit cell in Fig. 3. Segments A and B have length $d_A = d_B = 0.311d$ and the TL parameters are $L_{11} = L_{22} = L_0$, $C_{11} = 184.1C_0$, and $C_{22} = 3.5C_0$, where $L_0 = 1.256\,[\text{nH/m}]$ and $C_0 = 8.85\,[\text{pF/m}]$. Segment C has length $d_C = 0.378d$ and the TL parameters are $L_{11} = L_{22} = L_0$ and $C_{11} = C_{22} = C_0$. In these cases the TLs have no coupling inductance and capacitance, i.e., $L_{12} = L_{21} = C_{12} = C_{21} = 0$, within each segment. Coupling occurs only at interfaces between segments. All TLs have cutoff capacitance $C_c = 1\,[\text{pF.m}]$. The rotation angle $\delta\varphi_m$ in (13) is $45°$, and $d = 25$ mm corresponding to a DBE radian frequency $\omega_d \cong 2\pi\,(2.1\text{ GHz})$.

To generate the plots in Sec. IV are generated using the following parameters for a unit cell in Fig. 13: segment A has length $d_A = d/2 = 10$ mm and TL parameters $L_{11} = L_0$, $L_{22} = 0.95L_0$ and $C_{11} = C_0$, $C_{22} = 0.32C_0$ with no coupling inductance and capacitance, i.e, $L_{12} = L_{21} = C_{12} = C_{21} = 0$, where $L_0 = 0.451\,[\text{μH/m}]$ and $C_0 = 0.146\,[\text{nF/m}]$. Segment B has length $d_B = d_A$ and TL parameters $L_{11} = L_{22} = 1.2L_0$ and $C_{11} = C_{22} = 0.25C_0$ with mutual coupling parameters $L_{12} = L_{21} = 0.04L_0$ nH, and $C_{12} = C_{21} = 0.12C_0$, with all TLs having cutoff capacitance $C_c = 0.1\,[\text{pF.m}]$.

For the comparison in Sec. V, we use a 50 $[\Omega]$ TL to obtain the gain in Pierce type TWT with the same electrical length as that of the DBE structure described in the paragraph above. In the comparison of Fig. 16 (and only there) we assume in all cases distributed losses in the TLs described by the per-unit-length resistance of 0.01 $[\Omega/\text{m}]$ and a beam area of $A = 10^{-4}\ \text{m}^2$ corresponding to a normalized plasma frequency $\omega_p d/(2\pi c) = 0.013$.

## APPENDIX C.

## BLOCH IMPEDANCE OF THE "COLD" MTL PERIODIC STRUCTURE

The analysis of the Bloch impedance for a periodic 2-TL structure is more involved that the case of single periodic TL (for single period TL, it is formulated in a few textbooks such as in Chapter 8 in [55] ), for instance). We consider only the "cold" MTL structure, without interaction with the electron beam, therefore in this appendix (and only here) 4×1 vector state vectors are defined by $\mathbf{\Psi}(z) = \left[ \mathbf{V}^T(z) \quad \mathbf{I}^T(z) \right]^T$. The relative evolution equation is still $\partial_z \mathbf{\Psi}(z) = -j\underline{\mathbf{M}}(z)\mathbf{\Psi}(z)$  where  $\underline{\mathbf{M}}(z)$  is an  4×4  matrix representing only the cold structure, and the transfer matrix $\underline{\mathbf{T}}$, also an  4×4 matrix is translates the state vector across a unit cell as

$$\underline{\mathbf{T}}\,\mathbf{\Psi}(z) = e^{-jkd}\,\mathbf{\Psi}(z). \tag{C1}$$

We introduce the matrix $\underline{\underline{\mathbf{k}}}$ as a 2×2 diagonal matrix, whose diagonal elements are the Bloch wavenumbers with positive real values, i.e., $\underline{\underline{\mathbf{k}}} = \mathrm{diag}(k_1, k_2)$ where that the fact that $k_1, k_2, -k_1, -k_2$ are the four modal wavenumbers (eigenvalues) for the cold periodic structure holds for a reciprocal system (without the beam). We will also use $\underline{\mathbf{\Lambda}}$ as diagonal matrix whose diagonal elements are the eigenvalues of the (C1) via

$$\underline{\mathbf{\Lambda}} = \begin{pmatrix} e^{-j\underline{\underline{\mathbf{k}}}d} & \underline{\mathbf{0}} \\ \underline{\mathbf{0}} & e^{j\underline{\underline{\mathbf{k}}}d} \end{pmatrix}. \tag{C2}$$

Therefore it follows that the transfer matrix $\underline{\mathbf{T}}$ is written as $\underline{\mathbf{T}} = \underline{\mathbf{U}}\,\underline{\mathbf{\Lambda}}\,\underline{\mathbf{U}}^{-1}$ , where $\underline{\mathbf{U}}$ is a non-singular similarity transformation that diagonalize $\underline{\mathbf{T}}$, and is computed using the four eigenvectors of $\underline{\mathbf{T}}$ as $\underline{\mathbf{U}} = [\mathbf{\Psi}_1 \quad \mathbf{\Psi}_2 \quad \mathbf{\Psi}_3 \quad \mathbf{\Psi}_4]$ . At the exceptional point where DBE is manifested, $\underline{\mathbf{T}}$ is no longer diagonalizable; it becomes similar to a Jordan Block such that $\underline{\mathbf{\Lambda}} = \underline{\mathbf{J}}$ and

$$\underline{\mathbf{J}} = \begin{pmatrix} -1 & 1 & 0 & 0 \\ 0 & -1 & 1 & 0 \\ 0 & 0 & -1 & 1 \\ 0 & 0 & 0 & -1 \end{pmatrix}, \tag{C3}$$

while  a non-singular transformation $\underline{\mathbf{U}}$ that can be only defined using generalized eigenvectors [33], [51]. The following procedure is derived strictly when the $\underline{\mathbf{T}}$ is similar to a diagonal matrix $\underline{\mathbf{\Lambda}}$ , i.e., for all frequencies except at the DBE point.

We now introduce a basis-rotation non-singular matrix $\underline{\mathbf{P}}$ that transform the state vector $\mathbf{\Psi}$ into its modal representation $\mathbf{\Psi}'$ (a technique also referred to as "decoupling" the MTLs equations  [50], [53], [60]) using the transformation

$$\mathbf{\Psi}(z) = \begin{pmatrix} \underline{\underline{\mathbf{P}}}_v & \underline{\underline{\mathbf{0}}} \\ \underline{\underline{\mathbf{0}}} & \underline{\underline{\mathbf{P}}}_i \end{pmatrix} \mathbf{\Psi}'(z) = \underline{\mathbf{P}}\mathbf{\Psi}'(z), \quad \mathbf{V}(z) = \underline{\underline{\mathbf{P}}}_v \mathbf{V}'(z), \quad \text{and} \quad \mathbf{I}(z) = \underline{\underline{\mathbf{P}}}_i \mathbf{I}'(z), \tag{C4}$$

where modal voltage and currents relate to the physical voltage and current through $\mathbf{V}(z) = \underline{\underline{\mathbf{P}}}_v \mathbf{V}'(z)$ and $\mathbf{I}(z) = \underline{\underline{\mathbf{P}}}_i \mathbf{I}'(z)$ , with $\underline{\underline{\mathbf{P}}}_v$ and $\underline{\underline{\mathbf{P}}}_i$ being 2×2 non-singular matrices [49].  By using (C4) we can rewrite the (C1) as

$$\underline{\mathbf{T}}\,\underline{\mathbf{P}}\mathbf{\Psi}'(z) = e^{-jkd}\,\underline{\mathbf{P}}\mathbf{\Psi}'(z). \tag{C5}$$

Let us also define a new modal transfer matrix $\underline{\underline{\mathbf{\Pi}}}$ that is similar to $\underline{\mathbf{T}}$ and can be written as $\underline{\underline{\mathbf{\Pi}}} = \underline{\mathbf{P}}^{-1}\underline{\mathbf{T}}\underline{\mathbf{P}}$ then we rewrite the eigen-system in (C5) as $\underline{\underline{\mathbf{\Pi}}}\mathbf{\Psi}'(z) = e^{-jkd}\mathbf{\Psi}'(z)$. Afterwards, we define a diagonal (modal) Bloch impedance relating the modal voltage and current vectors as $\mathbf{V}'(z) = \underline{\underline{\mathbf{Z}}}_{D\pm}^{B}\mathbf{I}'(z)$ with the $\pm$ denoting the $\pm$ z-direction, $\underline{\underline{\mathbf{Z}}}_{D+}^{B} = \mathrm{diag}(Z_{1+}^{B}, Z_{2+}^{B})$. We associate $\underline{\underline{\mathbf{Z}}}_{D+}^{B}$ with $\underline{\underline{\mathbf{k}}}$ (modes propagating in the positive $z$-direction with $\mathrm{Re}(k_{1}), \mathrm{Re}(k_{2}) > 0$) and $\underline{\underline{\mathbf{Z}}}_{D-}^{B}$ with $-\underline{\underline{\mathbf{k}}}$ (modes propagating in the negative $z$-direction with $\mathrm{Re}(k_{1}), \mathrm{Re}(k_{2}) < 0$) for symmetry reasons in the reciprocal structure. The transfer matrix $\underline{\underline{\mathbf{\Pi}}}$ now translates $\mathbf{\Psi}'$ in the periodic structure across a unit cells and is written in terms of the a diagonal Bloch impedance and admittance $\underline{\underline{\mathbf{Z}}}_{D\pm}^{B}$ and $\underline{\underline{\mathbf{Y}}}_{D\pm}^{B} = \left(\underline{\underline{\mathbf{Z}}}_{D\pm}^{B}\right)^{-1}$, respectively [49], [53], as

$$\underline{\underline{\mathbf{\Pi}}} = \begin{pmatrix} \cos(\underline{\underline{\mathbf{k}}}d) & j\underline{\underline{\mathbf{Z}}}_{D+}^{B}\sin(\underline{\underline{\mathbf{k}}}d) \\ -j\underline{\underline{\mathbf{Y}}}_{D-}^{B}\sin(\underline{\underline{\mathbf{k}}}d) & \cos(\underline{\underline{\mathbf{k}}}d) \end{pmatrix}, \tag{C6}$$

which can be easily diagonalized, since it comprises "modal" or "decoupled" quantities in the form of diagonal matrices, namely $\underline{\underline{\mathbf{Z}}}_{D\pm}^{B}$ and $\underline{\underline{\mathbf{k}}}$. Also, we define $\cos(\underline{\underline{\mathbf{k}}}d) = \mathrm{diag}\left(\cos(k_{1}d), \cos(k_{2}d)\right)$. The diagonalization of $\underline{\underline{\mathbf{\Pi}}}$ is carried out to be in the form

$$\underline{\underline{\mathbf{\Pi}}} = \underline{\mathbf{W}} \begin{pmatrix} e^{-j\underline{\underline{\mathbf{k}}}d} & \underline{\underline{\mathbf{0}}} \\ \underline{\underline{\mathbf{0}}} & e^{j\underline{\underline{\mathbf{k}}}d} \end{pmatrix} \underline{\mathbf{W}}^{-1}, \quad \underline{\mathbf{W}} = \begin{pmatrix} \underline{\underline{\mathbf{1}}} & -j\sqrt{\underline{\underline{\mathbf{Z}}}_{D-}^{B}\,\underline{\underline{\mathbf{Z}}}_{D+}^{B}} \\ j\sqrt{\underline{\underline{\mathbf{Y}}}_{D-}^{B}\,\underline{\underline{\mathbf{Y}}}_{D+}^{B}} & \underline{\underline{\mathbf{1}}} \end{pmatrix}, \tag{C7}$$

where $\underline{\mathbf{W}}$ represents the eigenvector space of the modal transfer matrix $\underline{\underline{\mathbf{\Pi}}}$, and the square root is defined for a diagonal matrix as $\sqrt{\underline{\underline{\mathbf{Y}}}_{D-}^{B}} = \mathrm{diag}(+\sqrt{Y_{1-}^{B}}, +\sqrt{Y_{2-}^{B}})$. Notice the normalization in casting the matrix $\underline{\mathbf{W}}$ in the form (C7) so that the Bloch impedance/admittance can be explicitly identified. Moreover, note that $\underline{\mathbf{W}}$ is not uniquely defined (in the sense that multiplication of $\underline{\mathbf{W}}$ by an arbitrary constant will not change the eigenvectors space), however the Bloch impedance is uniquely defined [49], [53]. It follows from (C5) that the transfer matrix $\underline{\mathbf{T}}$ is written as

$$\underline{\mathbf{T}} = \underline{\mathbf{P}}\,\underline{\underline{\mathbf{\Pi}}}\,\underline{\mathbf{P}}^{-1} = \underline{\mathbf{P}}\,\underline{\mathbf{W}}\,\underline{\mathbf{\Lambda}}\,\underline{\mathbf{W}}^{-1}\,\underline{\mathbf{P}}^{-1} = \underline{\mathbf{U}}\,\underline{\mathbf{\Lambda}}\,\underline{\mathbf{U}}^{-1}, \tag{C8}$$

where we obtain $\underline{\mathbf{U}}$ using $\underline{\mathbf{P}}$ and $\underline{\underline{\mathbf{Z}}}_{D\pm}^{B}$ as

$$\underline{\mathbf{U}} = \underline{\underline{\mathbf{P}}}\underline{\underline{\mathbf{W}}} = \begin{pmatrix} \underline{\underline{\mathbf{P}}}_v & -j\underline{\underline{\mathbf{P}}}_v\sqrt{\underline{\underline{\mathbf{Z}}}^B_{D-}\ \underline{\underline{\mathbf{Z}}}^B_{D+}} \\ j\underline{\underline{\mathbf{P}}}_i\sqrt{\underline{\underline{\mathbf{Y}}}^B_{D-}\ \underline{\underline{\mathbf{Y}}}^B_{D+}} & \underline{\underline{\mathbf{P}}}_i \end{pmatrix}. \tag{C9}$$

To calculate the Bloch impedance as plotted in Fig. 5, we first evaluate the transformation $\underline{\mathbf{U}}$ numerically by computing the eigenvector space of the transfer matrix $\underline{\underline{\mathbf{T}}}$ from (C1) (whose constituent MTL parameters are described in Appendix B). Afterward, by equating $\underline{\mathbf{U}}$ to the matrix in (C9) we find the matrix $\underline{\underline{\mathbf{P}}}$ and Bloch impedances $\underline{\underline{\mathbf{Z}}}^B_{D\pm}$. Notice that this technique assumes that the Bloch impedance can be written as a diagonal matrix. At DBE, $\underline{\mathbf{U}}$ becomes singular at the DBE therefore the Bloch impedance is no longer diagonal. This technique however still valid around the close vicinity of DBE.